\documentclass[]{pasj02}
\usepackage[switch,mathlines]{lineno}
\usepackage{url}

\jyear{2024}
\Received{}
\Accepted{}

\begin{document} 

\title{Cloud-cloud collisions in the Antennae galaxies: Does high-speed collision suppress star formation?}

\author{
Shin \textsc{Inoue},\altaffilmark{1}\altemailmark\orcid{0009-0004-6347-0613}\email{sinoue@kusastro.kyoto-u.ac.jp}
Kouji \textsc{Ohta},\altaffilmark{1}\orcid{0000-0003-3844-1517}
Fumiya \textsc{Maeda}\altaffilmark{2}\orcid{0000-0002-8868-1255}
}
\altaffiltext{1}{Department of Astronomy, Kyoto University, Kitashirakawa-Oiwake-Cho, Sakyo-ku, Kyoto, Kyoto 606-8502, Japan}
\altaffiltext{2}{Research Center for Physics and Mathematics, Osaka Electro-Communication University, 18-8 Hatsucho, Neyagawa, Osaka 572-8530, Japan}




\maketitle

\begin{abstract}
Cloud-cloud collision (CCC) has been proposed as a mechanism for triggering massive star formation.
Observations in the Milky Way and nearby galaxies have revealed the presence of CCCs with collision velocity ($v_{\mathrm{col}}$) of $1$--$40$\>km\>s$^{-1}$, and the connection between star formation activity and the properties of colliding clouds has been investigated.
In this study, we expand the study to much faster ($\sim100$\>km\>s$^{-1}$) CCCs in a nearby colliding galaxies system, the Antennae galaxies.
We examine how star formation rate (SFR) on a sub-kpc scale depends on the $v_{\mathrm{col}}$ and mass ($M_{\mathrm{mol}}$) of giant molecular clouds (GMCs) across the Antennae galaxies, which show diverse star formation activity.
Furthermore, to examine the star formation process at a more fundamental level, we also investigate how the star formation efficiency (SFE) of a colliding GMC depends on its $v_{\mathrm{col}}$ and $M_{\mathrm{mol}}$. 
SFR is calculated using H$\alpha$ and mid-infrared data. From $\sim2000$ GMCs identified in the CO(1--0) data cube using the ALMA archival data, collision velocities are estimated based on the velocity dispersion among GMCs in a sub-kpc scale region, assuming random motion in three-dimensional space.
GMCs are considered to be colliding at a velocity of $\sim10$--$150$\>km\>s$^{-1}$.
We find that regions where high-speed collisions ($v_{\mathrm{col}}\sim100$\>km\>s$^{-1}$) of massive ($M_{\mathrm{mol}}\sim10^{7\mathchar`-\mathchar`-8}$\>$M_\odot$) GMCs are seen show the highest surface density of SFR.
Particularly, in the region with $v_{\mathrm{col}}\sim100$\>km\>s$^{-1}$, we find that SFR on a sub-kpc scale increases with increasing $M_{\mathrm{mol}}$
in the range of $\sim10^{6}$--$10^{8}$\>$M_\odot$.
The SFE of a colliding cloud is estimated to be 0.1\%--3.0\% without clear $M_{\mathrm{mol}}$ dependence, and the SFE is the lowest at the $v_{\mathrm{col}}\sim100$--$150$\>km\>s$^{-1}$.
These results suggest that the most active star formation in the Antennae galaxies seems to occur due to large GMC mass.
\end{abstract}


\section{Introduction}
\label{sec:introduction}
Gas compressin by cloud-cloud collision (CCC) has been discussed as one of the important processes to trigger formation of massive stars \citep{Oort46, Stone70a, Stone70b, Loren76, Gilden84, Scoville+86, HabeOhta92}.
Numerical simulations show that a shock compression induced by CCC produces high-mass pre-star cores of $\sim100$\>$M_\odot$ \citep{InoueFukui13, Takahira+14}.
In fact, recent observational studies of molecular gas in star-forming regions in the Milky Way, the Large Magellanic Cloud, the Small Magellanic Cloud, and nearby galaxies revealed evidences for molecular clouds of $\sim10^{3\mathchar`-\mathchar`-6}$\>$M_\odot$ are colliding at a relative velocity of $1$--$40$\>km\>s$^{-1}$, and massive stars form in the compressed regions (see \cite{Fukui+21} for a review).

Recent observational studies, however, suggest that high-speed CCC suppresses the formation of massive stars in nearby disk galaxies.
\citet{Maeda+21} studied connections between a diversity of star formation activities across galactic structures and collision velocity among giant molecular clouds (GMCs) in the nearby barred spiral galaxy NGC\,1300. In the galaxy, active massive star formation is seen in the arm region, while it is nearly absent in the bar region.
They suggest that GMCs collide at a higher velocity in the bar region ($\sim20$\>km\>s$^{-1}$) compared to the arm region ($\sim11$\>km\>s$^{-1}$ ).
They concluded that the high-speed CCC may suppress massive star formation in the bar region, resulting in differences in star formation activities between the arm and bar regions.
Numerical simulations of CCC \citep{Takahira+14, Takahira+18, Sakre+23} support this interpretation by showing that in case of high-speed collision, a collision duration is too short to form massive stars, resulting in the absence of them.
Similar trends are also seen in 18 nearby barred galaxies \citep{Maeda+23} and in the Milky Way \citep{Enokiya+21}.

\begin{figure}
 \begin{center}
 \includegraphics[width=80mm]{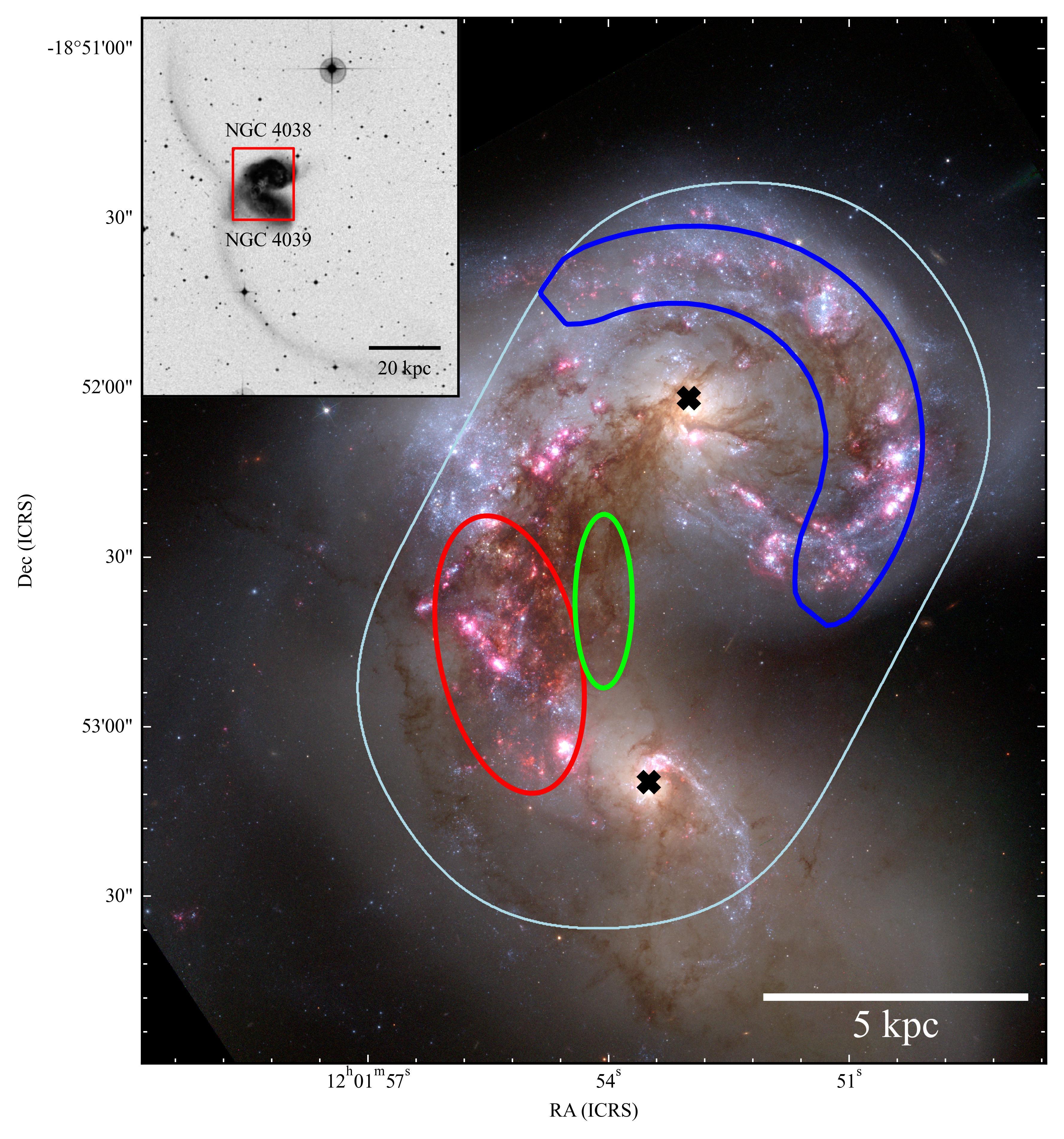}
 \end{center}
\caption{
Upper-left panel:
panoramic view of the Antennae galaxies (NGC\,4038 and NGC\,4039) in blue band of the Digital Sky Survey 1.
The red box indicates the FoV of the main panel.
Main panel: 
Zoom-up optical image of the central part of the Antennae galaxies taken with Hubble Space Telescope.
F435W, F550M, and a combination of the F814W and H$\alpha$ images are shown in blue, green, and red, respectively.
Black crosses mark the nuclei of NGC\,4038 and NGC\,4039, located at (RA, Dec)$_{\rm J2000.0}$ = (\timeform{12h01m53.0s}, \timeform{-18D52'01.9''}) and (\timeform{12h01m53.5s}, \timeform{-18D53'09.8''}), respectively, measured from radio continuum \citep{Zhang+01}.
The light blue solid line indicates the FoV of ALMA CO(1--0) observations.
The active overlap, non-active overlap, and arm-like regions are approximately indicated with red ellipse, green ellipse, and blue line, respectively (see text).
{
Alt text: A color photograph with a monochrome photograph in the top-left corner.
}
}
 \label{fig:Antennae_overview}
\end{figure}

In early-mid stage interacting systems,
the galaxies collide at around $100$--$400$\>km\>s$^{-1}$ 
(e.g., \cite{Gallagher+81}),
suggesting that GMCs also collide at speeds higher than the ‘high-speed’ of $20$\>km\>s$^{-1}$ seen in disk galaxies. Nevertheless, some systems exhibit active massive star formation, suggesting that even higher-speed collision can trigger the massive star formation. 
A good example is the Antennae galaxies (NGC\,4038/4039; Figure \ref{fig:Antennae_overview}).
The Antennae galaxies are nearby (22\>Mpc; \cite{Schweizer+08}) major-merger galaxies between two similar-mass disk galaxies (e.g., \cite{ToomreToomre72}).
The system consists of two main galaxies (NGC\,4038 \& 4039) with long tidal tails extending to north and south (upper left corner panel in figure \ref{fig:Antennae_overview}).
Numerical simulations claim that the collision started 100--600 Myr ago and they are now in around the second pericenter passage \citep{Karl+10, Privon+13, Renaud+15}. 
In the region indicated by red ellipse in figure \ref{fig:Antennae_overview} (refer to `active overlap region'), obscured active star formation and CCCs with collision velocities of $\sim100$\>km\>s$^{-1}$ were found \citep{Tsuge+21a, Tsuge+21b}.

Why is the massive star formation induced despite the higher-speed collisions? The key parameter would be the mass of GMCs. In the active overlap region, molecular gas is very abundant and supergiant molecular complexes (SGMCs) exist \citep{Wilson+00}.
\citet{Tsuge+21a, Tsuge+21b} reported that the masses of the GMCs colliding at $\sim100$\>km\>s$^{-1}$ are $\sim10^{7\mathchar`-\mathchar`-8}$\>$M_\odot$, which are significantly larger than the GMC mass in the barred galaxy NGC\,1300 ($\sim10^6$ $M_{\odot}$; \cite{Maeda+21}).
These findings suggest that if the GMC mass is large, even high-speed CCC triggers the formation of massive stars. Numerical simulations of CCC \citep{Takahira+18} support this idea by showing that the larger mass leads to an increase in size and/or density, resulting in a longer duration and/or higher gas accretion rate, thereby inducing massive star formation. 
However, the observational relationship between GMC mass and star formation activity in high-speed CCCs remains unclear, because the previous studies examined only a few cases of CCCs.
In the early universe, galaxy collisions are expected to occur frequently (e.g., \cite{LaceyCole93, Hopkins+08}), which has a significant impact on galaxy evolution.
Therefore, investigating the parameter dependence of CCCs in colliding galaxies is highly important for the understanding of the diverse aspects of massive star formation in galaxies.

In this study, we focus on the Antennae galaxies and investigate the relationship between star formation activity, collision velocity, and GMC mass across the wide region shown in figure \ref{fig:Antennae_overview}, where the diverse star-formation activities is seen.
In addition to the active overlap region (red ellipse in figure \ref{fig:Antennae_overview}), active star formation is observed in a western to northern arm-like region of NGC\,4038 (blue line in figure \ref{fig:Antennae_overview}).
In contrast to them, in the region indicated by green ellipse in figure \ref{fig:Antennae_overview}, no noticeable H\,\emissiontype{II} region exists though a dense dust lane extending north to south is seen (refer to `non-active overlap region').
This wide range of star formation activities across the Antennae galaxies ($\Sigma_{\rm SFR} \sim0.01$--$1$\>$M_\odot$\>yr$^{-1}$\>kpc$^{-2}$; see Section 3) may result from variations in collision velocity and cloud mass of CCCs.

Using sensitive high-resolution ($\sim10$--$50$\>pc) CO data, \citet{Tsuge+21a, Tsuge+21b} identified sites of high-speed CCC in the active overlap region, based on the characteristic observational signatures of CCC: Bridge features between the two clouds and complementary spatial distribution (e.g., \cite{HabeOhta92, Takahira+14, Fukui+21}).
It is challenging, however, to detect such characteristic signatures across the main two galaxies including in the arm-like region and the non-active overlap region. This difficulty arises because GMCs in these regions are expected to be smaller in both mass and size compared to those in the active overlap region.
Therefore, CO observations with much higher sensitivity and spatial resolution than those by \citet{Tsuge+21a, Tsuge+21b} would be necessary for identifying the CCC signatures.
Unfortunately, the resolution and sensitivity of the data we use in this study are not good enough to identify individual colliding cloud pairs.
Thus, we take a rather statistical approach similar to the method adopted by \citet{Maeda+21} using 70\>pc resolution data that allows us to identify GMC.

This paper aims to investigate how 
star formation rate (SFR) on a sub-kpc scale and star formation efficiency (SFE) of a colliding cloud depend on the collision velocity and the mass of the GMC by targeting the central part of the Antennae galaxies (figure \ref{fig:Antennae_overview}).
We estimate the collision velocity based on the velocity dispersion among the clouds as like \citet{Maeda+21}.
Section \ref{sec:data} describes the data sets we used.
Sections \ref{sec:SFRderivation} and \ref{sec:GMCidentification} show derivation of SFR and identification of the GMCs, respectively.
In Section \ref{sec:aperture_analysis}, we describe how to derive the collision velocity in detail.
The resultant relationship between SFR on a sub-kpc scale, collision velocity of GMCs, and GMC mass is shown in section \ref{sec:results}.
In section \ref{sec:discussion}, we discuss star formation efficiency for a colliding cloud.
Finally, section 8 presents a summary of this study.

\section{Data}
\label{sec:data}
\subsection{ $^{12}$CO(1--0) }
We made a $^{12}$CO(1--0) data cube using the ALMA Band 3 (115 GHz) archival data obtained under Cycle 6 project 2018.1.00272.S (PI. C. Wilson).
The observations were carried out with compact (C43-3) and extended (C43-5) configurations of 12 m array, as well as Atacama Compact Array (ACA; 7 m array and Total Power (TP) array).
The field-of-view (FoV) is indicated by the light blue solid line in figure \ref{fig:Antennae_overview}, which covers not only the active overlap region (red ellipse) but also the non-active overlap region (green ellipse), the western to northern star-forming arm-like region (blue line), and the nuclear region of the NGC\,4038 and NGC\,4039.
This FoV was observed with 13 and 5 pointing mosaics of 12 m and 7 m arrays, respectively.

The total on-source time over the FoV is, 2.5, 9.9, 7.6, and 9.3 hours, for the compact configuration of 12 m array, the extended configuration of 12 m array, 7 m array, and TP array, respectively.
For both the 12 m and 7 m arrays data, we utilized calibrated visibility data provided by the East Asian ALMA Regional Center. 
For the Total Power data, we used the image cube provided by the observatory.

We used the Common Astronomy Software Application (CASA; \cite{McMullin+07}) version 6.5.2. to obtain a science data cube that combined 12 m array, 7 m array and TP array.
Because the radio continuum is detected in the actively star-forming region (e.g., \cite{He+22}), we subtracted the continuum emission from the visibility data using \texttt{uvcontsub} task.
Then, we concatenated the visibility data sets of all the pointings of the 12 m and 7 m arrays into a single visibility data and conducted mosaicing and imaging using the \texttt{tclean} task to obtain a dirty cube applying a Briggs weighting with robust = 0.5.

Deconvolution was performed with a combination of multiscale CLEAN \citep{Cornwell08} and single-scale CLEAN \citep{Hogbom74} as a similar process of PHANGS-ALMA pipeline \citep{Leroy+21}.
First, we carried out a multiscale CLEAN 
until the maximum residual has $S/N$ of 3, using multiscale CLEAN parameter of $0''$, $1''$, $2''.5$, $5''$, $7''.5$, $10''$, $20''$, and $30''$.
In this multiscale CLEAN deconvolution, we used clean masks automatically created by automasking algorithm \citep{Kepley+20} incorporated into \texttt{tclean} task.
Second, we carried out a single CLEAN until the maximum residual has $S/N$ of 1.
Finally, we combined the TP cube using a feathering algorithm \citep{Cotton17} to create the final cube.

We set a pixel scale of $0.''13$ and binned two velocity channels into one channel ($\sim5.1$\>km\>s$^{-1}$).
The resulting synthesized beam size is $0''.72\times0''.56$ which corresponds to a physical scale of 76\>pc\,$\times$\,60\>pc at 22\>Mpc, with a position angle of $-87^\circ.3$.
The rms noise is $\sim0.63$\>mJy\>beam$^{-1}$, corresponding to $\sim0.14$\>K.
Finally, we corrected the primary beam attenuation.
We used the region within the primary beam correction factor smaller than 2.0, corresponding to the light blue boundary in figure \ref{fig:Antennae_overview}.

\subsection{Continuum-subtracted H$\alpha$}
We used a continuum-subtracted H$\alpha$ flux map of the Antennae galaxies created by \citet{Weilbacher+18} to derive the SFR.
They observed the Antennae galaxies using the wide-field mode ($1' \times 1'$) of Multi Unit Spectroscopic Explorer (MUSE; \cite{Bacon12}) on the Very Large Telescope (VLT).
A spectral range of $\sim4600$--$9350$\,\AA was covered with a resolution of $R\sim3000$, which corresponds to a velocity resolution of $\sim100$\>km\>s$^{-1}$.
They used MUSE pipeline \citep{Weilbacher+20} to obtain data cube with a sampling scale of $0.''2\times0.''2\times1.25$\,\AA.
They observed almost the same region shown in the main panel of figure \ref{fig:Antennae_overview} by 11 fields mosaicing under a typical seeing size of $\mathrm{FWHM}=0''.59$--$0''.76$.
They created a continuum-subtracted H$\alpha$ flux map from the data cube by fitting spectral region around H$\alpha$ emission.
The typical $1\sigma$ noise of the continuum-subtracted H$\alpha$ flux map is $3\times10^{-20}$\>erg\>s$^{-1}$\>cm$^{-2}$\>spaxel$^{-1}$.

\subsection{Far ultraviolet (FUV)}
As another star formation tracer, we used archival data of Galaxy Evolution Explorer (GALEX; \cite{Martin+05}) FUV image.
We downloaded calibrated image and sky background image from Mikulski Archive for Space Telescopes (MAST) \footnote{All the GALEX data used in this paper can be found in MAST at $\langle$\url{http://dx.doi.org/10.17909/4nhk-mm66}$\rangle$.
}.
We subtracted the sky background image from the calibrated image.
We converted the unit of the image from counts\>s$^{-1}$ to intensity using the conversion factor given by \citet{Leroy+19}.
We also corrected for the Galactic extinction using $A_{\mathrm{FUV}}=0.41$\>mag estimated by \citet{Leroy+19}.
The image shows an angular resolution of FWHM $\sim4''.2$ with a pixel scale of $1.''5$, and a typical noise estimated in a blank sky is $1\sigma=1.5\times10^{-3}$\>MJy\>sr$^{-1}$.

\subsection{Infrared}
The H$\alpha$ and FUV emission are expected to be affected by dust attenuation in the Antennae galaxies \citep{Klaas+10}.
In order to estimate the obscured SFR, we used 24\>$\micron$ data of the Multiband Imaging Photometer for Spitzer (MIPS; \cite{Rieke+04}) on the Spitzer Space Telescope.
The 24\>$\micron$ flux density is expected to contain the infrared cirrus component (IR cirrus), which is the dust thermal emission heated by an old stellar population unrelated to ongoing star formation.
To estimate and subtract the IR cirrus, 5.8\>$\micron$ data taken with Infrared Array Camera (IRAC; \cite{Fazio+04}) on the Spitzer Space Telescope and 70\>$\micron$ and 160\>$\micron$ data with Photodetector Array Camera and Spectrograph (PACS; \cite{Poglitsch+10}) on the Herschel Space Observatory were used (section \ref{sec:SFRderivation}). 
We also used $K_s$-band (2.2\>$\micron$) data of Two Micron All Sky Survey (2MASS; \cite{Jarrett+03}) to subtract a stellar continuum in IRAC 5.8\>$\micron$ and MIPS 24\>$\micron$ flux.

For the MIPS 24\>$\micron$ data, we used a background subtracted image created by \citet{Maeda+24}.
They downloaded a calibrated Super-Mosaic image from the Spitzer Heritage Archive and subtracted the mode value of the blank sky as the sky background.
The image shows an angular resolution of FWHM $\sim6''$ with a pixel scale of $2.''45$ and a typical noise estimated in a blank sky is $1\sigma=(5$--$6)\times10^{-2}$\>MJy\>sr$^{-1}$.

For the IRAC 5.8\>$\micron$ data, we downloaded a calibrated Super-Mosaic image from the Spitzer Heritage Archive\footnote{\url{https://catcopy.ipac.caltech.edu/dois/doi.php?id=10.26131/IRSA543}}.
Because the downloaded image shows a gradient in the sky, we estimated and subtracted the sky background by fitting the blank sky with a linear plane.
The image shows an angular resolution of FWHM $\sim1.''88$ with a pixel scale of $0.''6$ and a typical rms noise estimated in the blank sky is $1\sigma=(1$--$2)\times10^{-2}$\>MJy\>sr$^{-1}$.

For the PACS 70 and 160\>$\micron$ data, we downloaded a calibrated image (Observation ID: 1342187836 and 1342187837) from Herschel science archive
and subtracted the mode value of the blank sky as a sky background.
They show an angular resolution of FWHM $\sim5.''5$ and $11.''3$ with a pixel scale of $1.''6$ and $3''.2$, respectively \citep{Klaas+10}.
Typical rms noises of the both images estimated in the blank sky are $1\sigma=(2$--$4)\times10^{-4}$\>Jy\>pix $^{-1}$.

For the 2MASS $K_s$-band data, we downloaded a calibrated and sky subtracted image from 2MASS Large Galaxy Atlas\footnote{\url{https://catcopy.ipac.caltech.edu/dois/doi.php?id=10.26131/IRSA122}}.
The image shows a seeing size of FWHM $\sim3''.0$ with a pixel scale of $1''.0$ and a typical noise is $1\sigma=7\times10^{-29}$\>erg\>s$^{-1}$\>cm$^{-2}$\>Hz$^{-1}$ pix$^{-1}$.

\section{Star formation rate}
\label{sec:SFRderivation}
\subsection{SFR derived from H$\alpha$, FUV and 24\>$\micron$ flux}
To derive SFR, we used H$\alpha$ and 24\>$\micron$ data to compensate for the dust obscuration of H$\alpha$ emission,
\begin{equation}
\label{eq:Calzetti07.Ha24}
L_{\mathrm{H\alpha, cor}} = L_{\mathrm{H\alpha, obs}} + (0.031\pm0.006) L({\mathrm{24\>\micron}}),
\end{equation}
where $L({\mathrm{24\>\micron}})$ refers to $\nu_{\mathrm{24\>\micron}}L_{\mathrm{24\>\micron}}$ \citep{Calzetti+07}.
Using the conversion factor obtained by \citet{Murphy+11}, the surface density of SFR is derived as 
\begin{equation}
\label{eq:Murphy11.HaSFR}
\left(
\frac{
\Sigma_{\mathrm{SFR_ {H\alpha}}}}
 {M_\odot\>\mathrm{yr}^{-1}\>\mathrm{kpc}^{-2}}
 \right)
 = 
5.37 \times 10^{-42} 
 \left(\frac{L_{\mathrm{H\alpha, cor}}}{\mathrm{erg\>s^{-1}}}\right)
 \frac{1}{(S/\mathrm{kpc^2})},
\end{equation}
where $S$ is the covered area for each pixel.

We also estimated the surface density of SFR using FUV and 24\>$\micron$ data as
\begin{equation}
\label{eq:SFRfromFUV24}
\left(
\frac{
\Sigma_{\mathrm{SFR_ {FUV}}}}
 {M_\odot\>\mathrm{yr}^{-1}\>\mathrm{kpc}^{-2}}
 \right)
  = 
  \left(
  \frac{8.1\times10^{-2}I_{\mathrm{FUV}} + 3.2^{+1.2}_{-0.7}\times10^{-3}I_{\mathrm{24\>\micron}}}{\mathrm{MJy}\>\mathrm{sr}^{-1}}
  \right)
\end{equation}
\citep{Leroy+08}. 

When calculating SFR, we convolved MUSE H$\alpha$ and GALEX FUV images using the kernel derived by \citet{Aniano+11} to match the angular resolution of MIPS 24\>$\micron$ data and regridded the images to have the same angular sampling as MIPS 24\>$\micron$ image.

\subsection{Estimation of IR cirrus contribution at 24\>$\micron$}
The MIPS 24\>$\micron$ flux density contains the IR cirrus component.
Especially, in an inactive star-forming region, the contamination is expected to be non-negligible.
Thus, we estimated and subtracted the cirrus contamination from the 24\>$\micron$ flux.
We employed two independent methods described below.

\subsubsection{Empirical method}
Spectral energy distribution (SED) of dust emission in a star-forming galaxy is composed of warm dust heated by newly formed massive stars and cold dust of IR cirrus (e.g., \cite{ Compiegne+11, Casey12}).
In the wavelength range of 3--8\>$\micron$, dust emission from Polycyclic Aromatic Hydrocarbon (PAH) dominates, and in the mid- to far-IR, thermal dust emission dominates.
The contribution from the warm dust is larger than that from the cold dust around 24\>$\micron$, allowing the intensity from the warm dust around $\sim24$\>$\micron$ to be used to estimate the obscured SFR using equations (\ref{eq:Calzetti07.Ha24})--(\ref{eq:SFRfromFUV24}).
In contrast, in the wavelength range of 3--8\>$\micron$, the contribution by warm dust is relatively small compared to that at $\sim24$\>$\micron$.
Furthermore, the shape of the IR cirrus SED is considered to be almost the same regardless of the radiation intensity of the old stellar population, and its luminosity changes almost linearly depending on the radiation intensity.
Therefore, the SED shape of the IR cirrus in the star-forming region can be estimated from the shape of the SED in a non-star-forming region, and the contribution of the IR cirrus at the star-forming region would be estimated by scaling their SED luminosity in 3--8\>$\micron$.
In this study, we used IRAC 5.8\>$\micron$ as a band to scale the luminosity of the IR cirrus SED.

At 5.8\>$\micron$ and 24\>$\micron$ flux, there is also contamination by stellar continuum.
To eliminate this contamination, we assumed that stellar continuum at wavelength longer than $\sim2$\>$\micron$ is approximated by the 5000 K black body (e.g., \cite{Munoz-Mateos+09, Ciesla+14}).
We scaled the 5000 K black body at 2.2\>$\micron$ intensity using 2MASS $K_s$-band image and subtracted it from 5.8\>$\micron$ and 24\>$\micron$ intensity as below,
\begin{equation}
I_{\mathrm{5.8\>\micron}}^{\mathrm{dust}} = I_{\mathrm{5.8\>\micron}} - 0.229 I_{\mathrm{2.2\>\micron}},
\end{equation}
\begin{equation}
I_{\mathrm{24\>\micron}}^{\mathrm{dust}} = I_{\mathrm{24\>\micron}} - 0.016 I_{\mathrm{2.2\>\micron}},
\end{equation}
where $I^{\mathrm{dust}}$ represents the intensity of dust emission.
The fractions of stellar continuum at 5.8\>$\micron$ and 24\>$\micron$ intensity are 10\%--40\% and $\sim2$\%, respectively, and this fraction changes less than 10\% if we use the 3000 K black body as the approximation of the stellar continuum.

\begin{figure}
 \begin{center}
  \includegraphics[width=80mm]{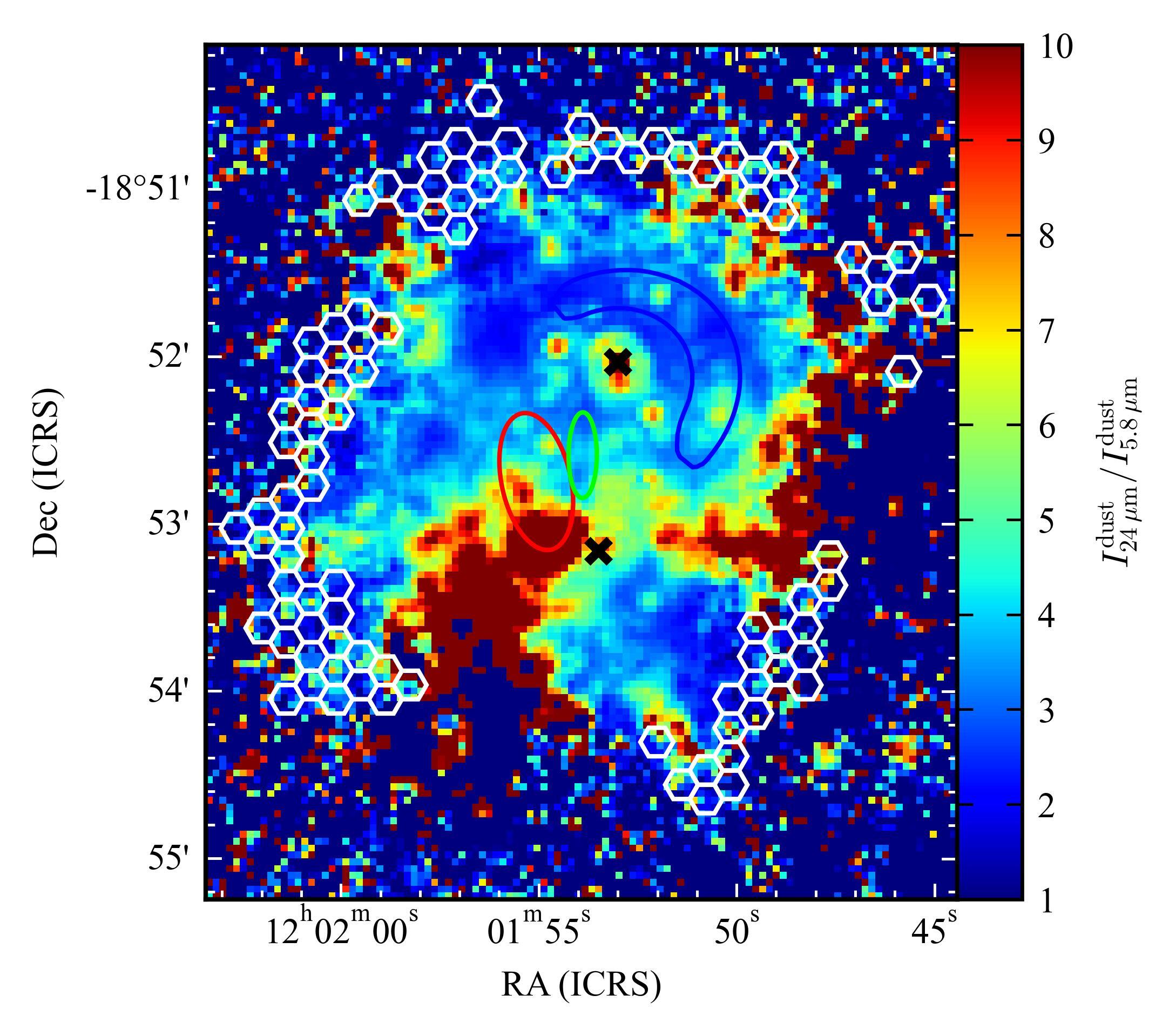}
 \end{center}
\caption{
Map of the intensity ratio of $I^{\mathrm{dust}}_{\mathrm{24\>\micron}}/I^{\mathrm{dust}}_{\mathrm{5.8\>\micron}}$ in the Antennae galaxies.
White hexagons are apertures in which we derived the typical value of $I^{\mathrm{dust}}_{\mathrm{24\>\micron}}/I^{\mathrm{dust}}_{\mathrm{5.8\>\micron}}$ to estimate the SED shape of the IR cirrus in the Antennae galaxies.
(See text.)
Black crosses indicate the nuclei of NGC\,4038 and NGC\,4039.
Red ellipse, green ellipse, and blue line are the same as those in figure \ref{fig:Antennae_overview}.
{Alt text: A map showing the ratio of the intensity of dust at 24 micron to 5.8 micron from 1 to 10.}
}
 \label{fig:24.58.cirrusaperture}
\end{figure}

Figure \ref{fig:24.58.cirrusaperture} shows the intensity ratio of $I^{\mathrm{dust}}_{\mathrm{24\>\micron}}/I^{\mathrm{dust}}_{\mathrm{5.8\>\micron}}$, which can be a proxy for the SED shape.
In the outer regions of the Antennae galaxies, $I^{\mathrm{dust}}_{\mathrm{24\>\micron}}/I^{\mathrm{dust}}_{\mathrm{5.8\>\micron}}$ is smaller than that in the inner region of the Antennae galaxies,
where the IR cirrus is dominant and star formation is negligible.
We set apertures in the outer region shown in figure \ref{fig:24.58.cirrusaperture} (white hexagon with $6''$ sides), 
and calculated $I^{\mathrm{dust}}_{\mathrm{24\>\micron}}/I^{\mathrm{dust}}_{\mathrm{5.8\>\micron}}$ in each aperture.
We estimated the typical intensity ratio of IR cirrus $\langle I^{\mathrm{dust}}_{\mathrm{24\>\micron}}/I^{\mathrm{dust}}_{\mathrm{5.8\>\micron}} \rangle _{\mathrm{cirrus}}$ as the mode value of the intensity ratio in the apertures.
The estimated value of $\langle I^{\mathrm{dust}}_{\mathrm{24\>\micron}}/I^{\mathrm{dust}}_{\mathrm{5.8\>\micron}} \rangle _{\mathrm{cirrus}}$ is $2.1 \pm 0.4$.
Using $\langle I^{\mathrm{dust}}_{\mathrm{24\>\micron}}/I^{\mathrm{dust}}_{\mathrm{5.8\>\micron}} \rangle _{\mathrm{cirrus}}$, the IR cirrus contribution at 24\>$\micron$ in a star forming region is estimated as 
\begin{equation}
\label{eq:24umcirrus}
I_{\mathrm{24\>\micron},~\mathrm{cirrus}} = \langle I_{\mathrm{24\>\micron}}^{\mathrm{dust}}/I_{\mathrm{5.8\>\micron}}^{\mathrm{dust}}\rangle_{\mathrm{cirrus}} \cdot I_{\mathrm{5.8\>\micron}}^{\mathrm{dust}}.
\end{equation}
Note that in this procedure, we matched the angular resolution of IRAC 5.8\>$\micron$ and 2MASS $K_s$-band images to MIPS 24\>$\micron$ image using the kernel
derived by \citet{Aniano+11} and regridded to have the same angular sampling as
MIPS 24\>$\micron$ image.

\subsubsection{Model based method}
We also estimated the IR cirrus contribution using the dust model by \citet{DraineLi07}.
\citet{DraineLi07} used a silicate-graphite-PAH dust model that reproduces the average Milky Way extinction curve and presented emission spectrum of dust heated by various starlight radiation fields.
They assumed that the starlight is the composition of the background radiation field ($U_{\mathrm{min}}$) representing old stellar population --- the emission of the dust heated by $U_{\mathrm{min}}$ corresponds to the IR cirrus --- and radiation field of power-law distribution of the starlight intensity representing newly formed stars.
They introduced a parameter of $\gamma$ as a dust mass fraction radiated by power-law component to reproduce various starlights.

\citet{DraineLi07} calculated the following two flux ratios with respect to $U_{\mathrm{min}}$ and $\gamma$ based on their model,
\begin{equation}
\label{eq:Draine07:P24}
P_{24} = \frac{\langle\nu F_\nu\rangle_{24}}{\langle\nu F_\nu\rangle_{71} + \langle\nu F_\nu\rangle_{160}},
\end{equation}
\begin{equation}
\label{eq:Draine07:R71}
R_{71} = \frac{\langle\nu F_\nu\rangle_{71}}{\langle\nu F_\nu\rangle_{160}},
\end{equation}
i.e., $U_{\mathrm{min}}$-$\gamma$ grid on $P_{24}$-$R_{71}$ plane.
Thus, by comparing this grid and observed values of $P_{24}$ and $R_{71}$, we can estimate $U_{\mathrm{min}}$ and $\gamma$.

We derived $P_{24}$ and $R_{71}$ using the observed data and compared it with the figure 21 by \citet{DraineLi07} to estimate $U_{\mathrm{min}}$ and $\gamma$.
Using the $U_{\mathrm{min}}$ and $\gamma$ thus obtained and the model spectrum, we calculated the 24\>$\micron$ intensities originated from the cirrus and star-formation and estimated the fraction of the IR cirrus at 24\>$\micron$.
\citet{DraineLi07} also introduced another parameter, PAH abundance.
However, this parameter does not change the $U_{\mathrm{min}}$ and $\gamma$ obtained significantly.
We fixed this parameter as a typical value (4.6\%) in the Milky Way.

In this estimation of the parameters and the IR cirrus contribution, we used fluxes of the MIPS 24\>$\micron$, PACS 70\>$\micron$ and PACS 160\>$\micron$ for $\langle\nu F_\nu\rangle_{24}$, $\langle\nu F_\nu\rangle_{71}$ and $\langle\nu F_\nu\rangle_{160}$, respectively.
Note again that we matched the angular resolution of all the images to PACS 160\>$\micron$ image using the kernel derived by \citet{Aniano+11} and regridded to have the same angular sampling as PACS 160\>$\micron$ image.

We compared this model-estimated fraction of the IR cirrus in 24\>$\micron$ flux with that of the empirically estimated values.
They agree with each other within a factor of 2.
The resulting IR cirrus contribution in the active overlap region 
is not large ($\sim10$\%--$20$\%), and that in the non-active overlap region and the arm-like region
is $\sim50$\%--$90$\%.
In the following discussion, we used the 24\>$\micron$ image after subtracting the empirically estimated IR cirrus component.
Note that, in the comparison of these two models, we match the angular resolution of the empirically estimated IR cirrus image to the model-estimated IR cirrus image using the kernel of \citet{Aniano+11} and regridded the images to have the same angular sampling as model-estimated IR cirrus image.

\subsection{SFR in the Antennae galaxies}
\begin{figure*}
 \begin{center}
  \includegraphics[width=160mm]{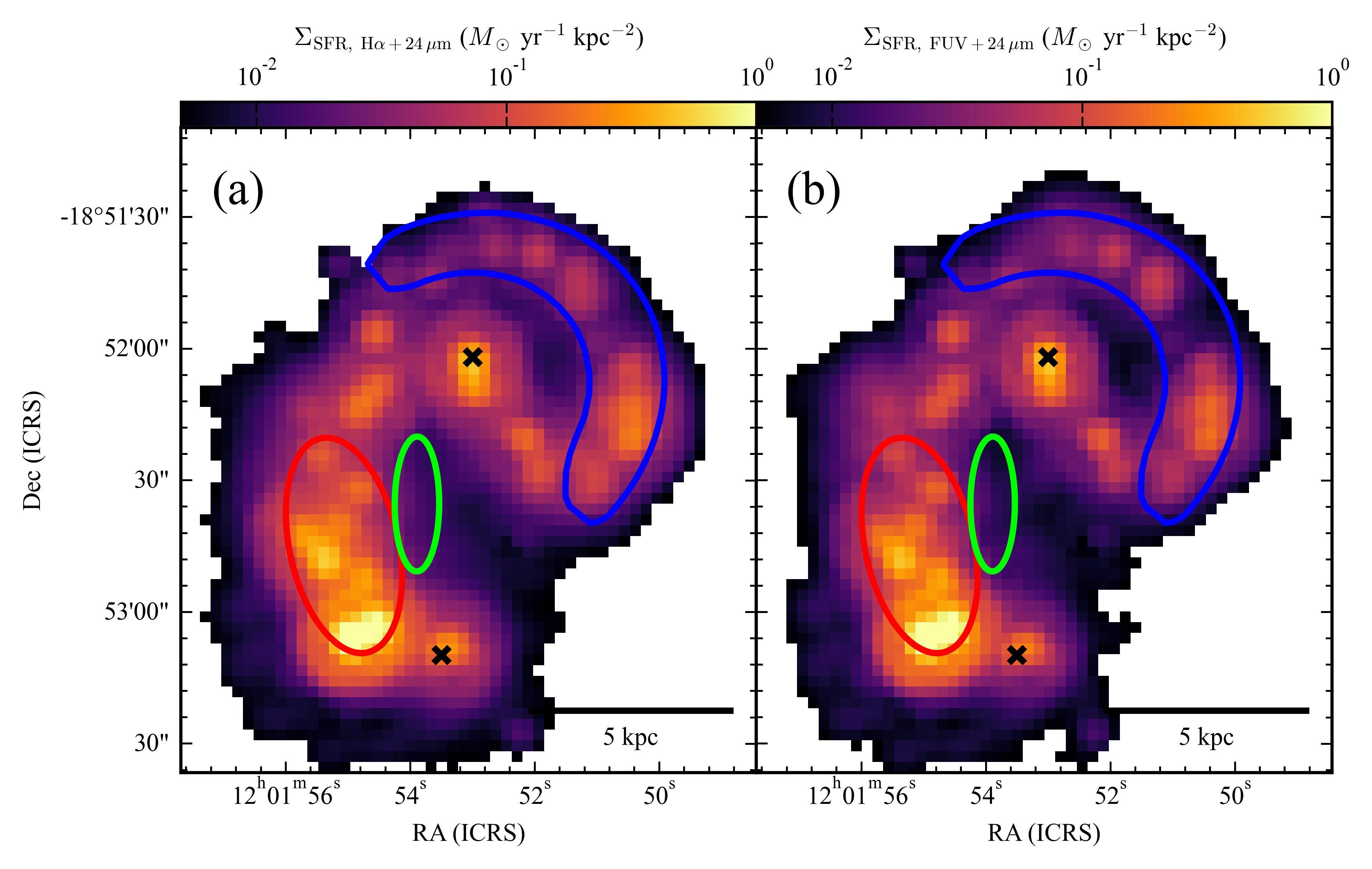}
 \end{center}
\caption{
Surface density of SFR in the Antennae galaxies.
Panel (a) and (b) shows the $\Sigma_{\mathrm{SFR}}$ derived from H$\alpha$ and 24\>$\micron$ data, and from FUV and 24\>$\micron$ data, respectively.
Black crosses indicate the nuclei of NGC\,4038 and NGC\,4039.
Red ellipse, green ellipse, and blue line are the same as those in figure \ref{fig:Antennae_overview}.
{Alt text: Two aligned maps labeled with a and b.
Two maps show the surface density of star formation rate derived from different data with a range from 0.01 to 0.1 Solar mass per year per square kiloparsec.}
}
 \label{fig:SFRmap.Ha24.FUV24}
\end{figure*}

Figure \ref{fig:SFRmap.Ha24.FUV24} shows the surface density of the SFR ($\Sigma_{\mathrm{SFR}}$) map of the Antennae galaxies derived from H$\alpha$ and 24\>$\micron$ flux (panel (a)) and from FUV and 24\>$\micron$ flux (panel (b))
with the angular resolution of $\sim6''=600$\>pc.
Antennae galaxies show a wide range of $\Sigma_{\mathrm{SFR}}\sim0.01$--$1$\>$M_\odot$\>yr$^{-1}$\>kpc$^{-2}$.
$\Sigma_{\mathrm{SFR}, ~\mathrm{H\alpha+24\>\micron}}$ and $\Sigma_{\mathrm{SFR}, ~\mathrm{FUV+24\>\micron}}$ at each position agrees well in the region of higher value ($\Sigma_{\mathrm{SFR}}\sim1$\>$M_\odot$\>yr$^{-1}$\>kpc$^{-2}$) and agrees with scatter within a factor of 2 even in the region of lower value ($\Sigma_{\mathrm{SFR}} \lesssim 0.1$\>$M_\odot$\>yr$^{-1}$\>kpc$^{-2}$).
In the following discussions, we use $\Sigma_{\mathrm{SFR}}$ derived from H$\alpha$ and 24\>$\micron$ flux.
Final conclusions of this paper do not change if we use $\Sigma_{\mathrm{SFR}}$ derived from FUV and 24\>$\micron$ flux.
The highest value of $\Sigma_{\mathrm{SFR}}\gtrsim 10^{-0.4}$\>$M_{\odot}$\>yr$^{-1}$\>kpc$^{-2}$ is seen in the active overlap region (red ellipse).
$\Sigma_{\mathrm{SFR}}$ in the arm-like region (blue line) in NGC\,4038 is $\sim 0.1$\>$M_{\odot}$\>yr$^{-1}$\>kpc$^{-2}$. 
$\Sigma_{\mathrm{SFR}}$ in the non-active overlap region (green ellipse) is low ($\lesssim10^{-1.3}$\>$M_\odot$\>yr$^{-1}$\>kpc$^{-2}$).
The uncertainty in $\Sigma_{\mathrm{SFR}}$, due to the propagation of uncertainty from the cirrus estimation, is relatively low in the regions with higher $\Sigma_{\mathrm{SFR}}$ values ($\sim 1$\>$M_\odot$\>yr$^{-1}$\>kpc$^{-2}$), but can be $\sim20$\%--$50$\% in the regions with lower values ($\sim 0.01$\>$M_\odot$\>yr$^{-1}$\>kpc$^{-2}$).

\section{GMCs identification and basic properties of GMCs}
\label{sec:GMCidentification}
\subsection{GMC identification}
We took two steps to identify GMCs in the Antennae galaxies.
First, following the method used by \citet{Rosolowsky+06}, we identified and masked significant emissions from the position--position--velocity (PPV) data cube of CO(1--0).
Second, 
we decomposed the significant emissions into individual GMCs by using 3D clump-finding algorithm PYCPROPS
\citep{Rosolowsky+21}.
Here, we briefly describe the GMC identification process and the algorithm of PYCPROPS.

For identifying and masking the significant emissions, we adopt the strategy presented by \citet{Rosolowsky+06}.
We first identified voxels with signal above $h\sigma$ ($\sigma$ is the rms noise of the data cube) in at least two adjacent velocity channels in the cube, then extended the area to include adjacent pixels above $l\sigma$.
We chose $h=5$ and $l=2$, and $\sigma=0.14$\>K.
Adopting $h$ of 4 or 6 and $l$ of 1.5 or 2.5 results in less than 6\% change of the total flux in the mask.

PYCPROPS is a publicly available Python package used to identify and characterize GMCs. 
It implements the CPROPS algorithm originally described by \citet{Rosolowsky+06} and takes an advantage of the fast dendrogram algorithm provided by the ASTRODENDRO package for the segmentation of emissions.
PYCPROPS is fully described by \citet{Rosolowsky+21}, thus we provide a brief summary.
PYCPROPS first identifies the local maxima in the significant emission masked as described above.
The local maxima are required to satisfy the following conditions:
\begin{enumerate}
    \item Its intensity ($T_\mathrm{max}$) must be at least $\delta\sigma$ higher than the merging level ($T_\mathrm{merge}$), which is defined as the highest value contour containing a given local maximum and one other neighbor.
    \item The number of pixels above the level of $T_\mathrm{merge}$ must be larger than the quarter of the beam size.
\end{enumerate}
In this study, we used $\delta = 2\sigma$. 
The choice of the parameter $\delta$ has the largest impact on the results.
However, if we use $\delta = 1.5\sigma$ or $2.5\sigma$, the final conclusions of this paper are not affected (see the Appendix).
In this study, each local maximum that satisfies the above conditions is assigned to an individual independent cloud (i.e., sigdiscont parameter is set to be 0.0).
Minimum spatial and velocity intervals between local maxima are not required (i.e., specfriends = 0 and friends = 0).

Finally, surrounding emission around the identified local maxima in the emission mask is assigned to the local maxima with Watershed algorithm \citep{vanderWalt+14}; boundaries of the GMCs are determined based on a {\it compactness parameter} \citep{NeubertProtzel14}.
The compactness parameter controls the smoothness of the boundary of the GMCs, with larger values resulting in smoother and more linear boundaries.
The appropriate value for this parameter needs to be determined based on the spatial resolution. 
When the resolution is low, since a GMC appears to be an ellipse with a smooth boundary similar to the beam shape, a larger value is appropriate.
However, if the resolution is high enough to resolve complex boundaries, a smaller value ($<1$) is preferable (see Appendix 1 by \cite{Demachi+24}).
The resolution of the CO image in this study is $\sim70$\>pc, which is higher than that of PHANGS-ALMA \citep{Rosolowsky+21} ($\sim 90$\>pc), and the CO distribution appears complex.
Therefore, we adopted a small compactness parameter of 0.001.
The difference in compactness parameter affects the mass and the size of the GMC, but even if we used 0.1 or 0.00001 and below as compactness parameter, the final conclusions are not affected (see the Appendix).

\subsection{Derivation of physical properties of the GMCs}
The identified GMCs comprise only the emission within the contour threshold of $2\sigma$.
PYCPROPS extrapolates boundaries of the GMCs to the expected 0 K contour threshold when measuring the physical properties of the GMCs as described in detail by \citet{Rosolowsky+06}.
To estimate the intrinsic size and line width, PYCPROPS also deconvolves the distribution of GMCs with a beam size and a channel width.

Line width ($\sigma_v$) is simply defined by the intensity-weighted second moment of the velocity but extrapolated and deconvolved with the channel width.

Size is derived as 
\begin{equation}
\label{eq:GMC.rad}
R = \eta\sqrt{\sigma_{\mathrm{maj}}\sigma_{\mathrm{min}}},
\end{equation}
where $\sigma_{\mathrm{maj}}$ and $\sigma_{\mathrm{min}}$ is the extrapolated and deconvolved intensity-weighted second moment of the positions along the major and minor axis, respectively.
By assuming that the distribution of the surface brightness of a GMC is a two-dimensional Gaussian, $\eta$ is calculated to be $\sqrt{2\ln 2} \sim 1.18$ \citep{Rosolowsky+21}.

The luminosity mass of a GMC is derived by converting the integrated flux in the GMC as 
\begin{equation}
\label{eq:GMC.Mmol}
M_{\mathrm{mol}} = \alpha_{\mathrm{CO}} \sum_i T_i \Delta v \Delta x \Delta y,
\end{equation}
where $\alpha_{\mathrm{CO}}$ is the CO-to-H$_2$ conversion factor in units of $M_\odot$\>(K\>km\>s$^{-1}$\>pc$^2$)$^{-1}$, $T_i$ is the brightness temperature of each pixel in the GMC in K, $\Delta x$, and $\Delta y$ are the pixel scale in pc, and $\Delta v$ is the channel width in $\rm km$\>s$^{-1}$.
The values of $\alpha_{\mathrm{CO}}$ in the Antennae galaxies are reported to range from $1.1$ to $6.54$\>$M_\odot$\>(K\>km\>s$^{-1}$\>pc$^2$)$^{-1}$ \citep{Wilson+03, Zhu+03, He+24} and show variation over the galactic structures \citep{He+24}.
In this study, we adopt a single value of $4.36$\>$M_\odot$(K\>km\>s$^{-1}$\>pc$^2$)$^{-1}$ as $\alpha_{\mathrm{CO}}$ for the entire region, which is approximately the midpoint of the range and corresponds to the typical value in the Milky Way (e.g., \cite{Bolatto+13}). 
Even if we use a different single value of $\alpha_{\mathrm{CO}}$, the final conclusions are not affected because this modification results in a same factor change in the GMC mass across all the regions.
As for the region-dependent values of $\alpha_{\mathrm{CO}}$, \citet{He+24} derived spatially resolved $\alpha_{\mathrm{CO}}$ in the Antennae galaxies at 150 pc scale.
Resolved $\alpha_{\mathrm{CO}}$ values are derived in the active overlap region, the nuclear regions, and a tip of the southern edge of the arm-like region (their figure 15).
$\alpha_{\mathrm{CO}}$ in these regions varies up to $\sim$60 \% except for the last region.
This does not affect the estimation of the GMC mass so much in this study.

\subsection{GMCs in the Antennae galaxies}
\begin{figure}
 \begin{center}
  \includegraphics[width=80mm]{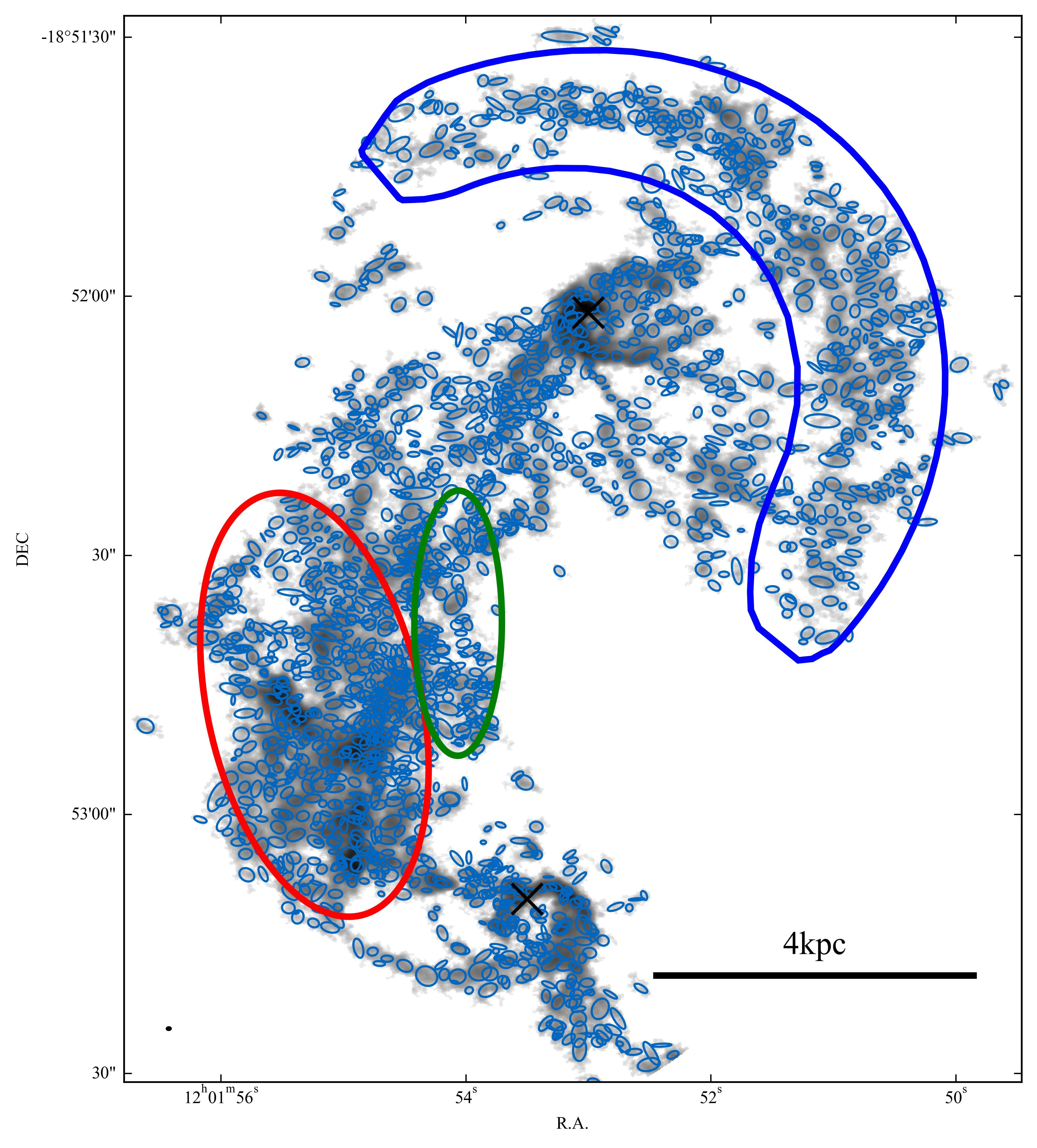}
 \end{center}
\caption{
GMC spatial distribution in the Antennae galaxies superimposed on the velocity integrated CO(1--0) map (gray-scale).
Blue ellipses show GMCs identified with the measured major, minor axes, and position angles.
Black crosses indicate the nuclei of NGC\,4038 and NGC\,4039.
Red ellipse, green ellipse, and blue line are the same as those in figure \ref{fig:Antennae_overview}.
{Alt text: An image of the central part of the Antennae galaxies with ellipses showing the distribution of the GMCs.}
}
 \label{fig:GMC_ellipse}
\end{figure}

\begin{figure*}
 \begin{center}
  \includegraphics[width=160mm]{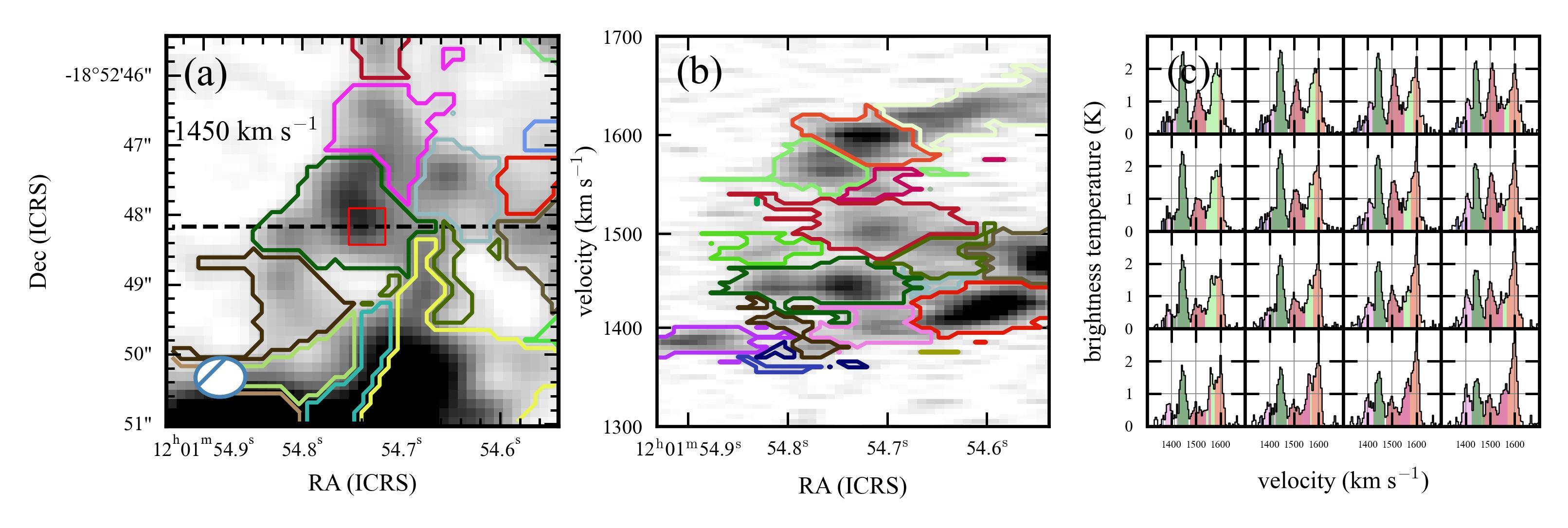}
 \end{center}
\caption{
(a) CO(1--0) channel map at a velocity of $\sim1450$\>km\>s$^{-1}$ (gray scale).
Colored contours show the boundaries of the GMCs identified.
(b) CO(1--0) position-velocity map along the black dashed line in panel (a).
The colored contours show the boundaries of the GMCs; the same color is used for the same GMC in panel (a).
(c) Velocity profiles in the region shown in the red square ($4 \times 4$ pixels) in panel (a).
The profiles are colored with the same color in panels (a) and (b) to indicate which components are assigned to the GMCs indicated in panels (a) and (b).
{Alt text: Two maps and a graph labeled from a to c.
In panel (b), the horizontal axis shows the position and the vertical axis shows the velocity.
In panel (c), the horizontal axis shows the velocity and the vertical axis shows the brightness temperature of the profiles.
}
}
 \label{fig:GMCidentification.example}
\end{figure*}

We identified 1759 GMCs in the Antennae galaxies.
Figure \ref{fig:GMC_ellipse} shows their spatial distribution represented by blue ellipses.
The distribution of the GMCs in the overlap and nuclear regions is complex, with many GMCs overlapping the same line-of-sight.
Figure \ref{fig:GMCidentification.example} shows an example of the identified GMCs in one of the most complicated regions.
Panel (a) displays the GMC boundaries in a channel map of the CO(1--0) at a velocity of $\sim1450$\>km\>s$^{-1}$.
Panel (b) shows the GMCs boundaries in a position--velocity diagram along the black dashed line in panel (a). 
Panel (c) shows velocity profiles in the region shown in the red square ($4 \times 4$ pixels) in panel (a).
The profiles are colored with the same color in panels (a) and (b) to indicate which parts of the profiles are assigned to each GMC.
Figure \ref{fig:GMCidentification.example} demonstrates that PYCPROPS reasonably works to separate the distribution of molecular gas into GMCs, even within the complex spatial and velocity distribution of the molecular gas; 
panels (a) and (b) show that each significant local maximum is identified as one cloud and the boundaries are drawn along the valleys between other peaks.
Panel (c) shows that each peak of the CO profile is assigned to a separate GMC.

\begin{figure*}
 \begin{center}
  \includegraphics[width=160mm]{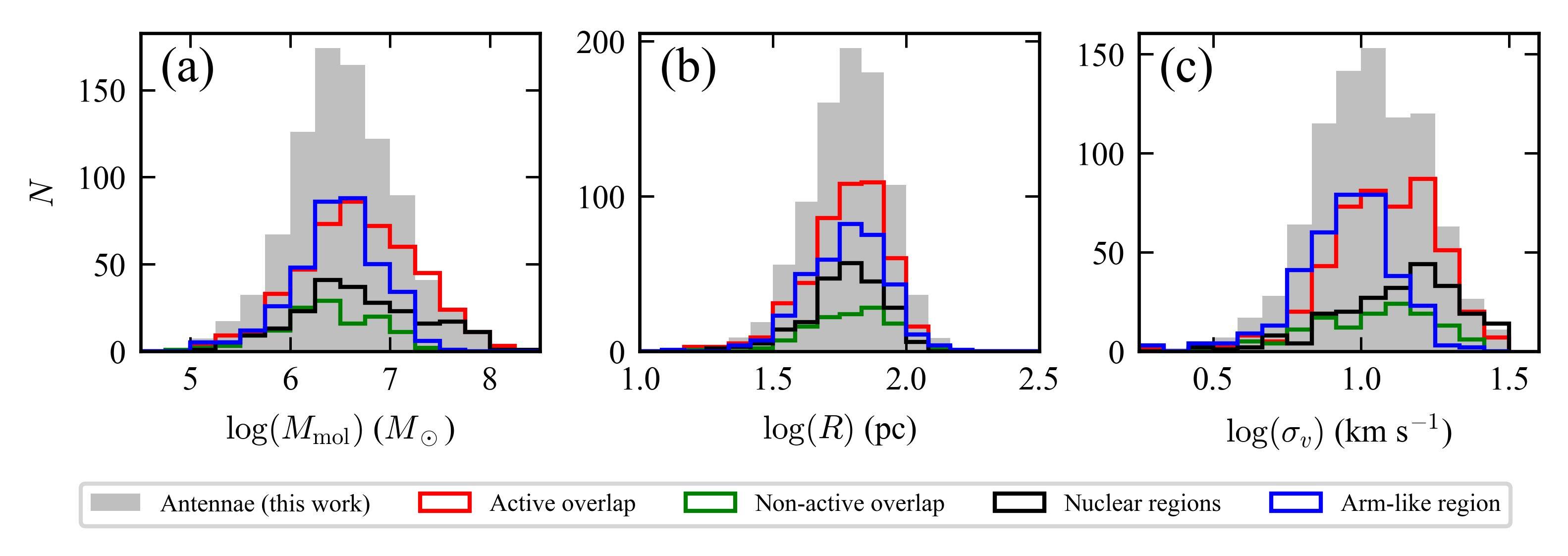}
 \end{center}
\caption{
Histogram of (a) mass ($M_{\mathrm{mol}}$), (b) radius ($R$), and (c) velocity dispersion ($\sigma_v$ ) of GMCs in the Antennae galaxies.
Gray filled histogram is for all the GMCs identified, but scaled down by a factor of 2 for better visualization.
Red, green, black, and blue histograms are for the GMCs in the active overlap, non-active overlap, nuclear, and arm-like regions, respectively (figure \ref{fig:Antennae_overview}).
The GMCs in the nuclear regions are those within 1\>kpc from the nuclei indicated in figure \ref{fig:Antennae_overview}.
Alt text: Three histograms labeled from a to c.
}
 \label{fig:GMC_general_properties}
\end{figure*}

Figure \ref{fig:GMC_general_properties}
shows the histogram of the properties of the GMCs.
The panel (a), (b), and (c) shows $M_{\mathrm{mol}}$, $R$, and $\sigma_v$, respectively, with
the median value of $M_{\mathrm{mol}}\sim 10^{6.5}$\>$M_\odot$, $R\sim60$\>pc, and $\sigma_v\sim10$\>km\>s$^{-1}$, respectively.
In the active overlap and nuclear regions, massive GMCs with $M_{\mathrm{mol}}\sim10^8$\>$M_\odot$ are identified.
Many of the GMCs in the Antenna galaxies show $\sigma_v$ larger than 10\>km\>s$^{-1}$.

Previous studies \citep{Ueda+12, Wei+12, ZaragozaCardiel+14} observed the molecular gas in the active overlap region and the nuclear regions in the Antennae galaxies with a $\sim60$--$100$\>pc resolution and identified GMCs.
They found a significant population of massive GMCs ($\sim10^8$\>$M_\odot$) in the active overlap and nuclear regions, which are usually not seen in normal spiral galaxies and our results are consistent with these results.
These previous studies \citep{Ueda+12, Wei+12, ZaragozaCardiel+14} also reported that the velocity dispersions of the GMCs in the Antennae galaxies of $10$--$30$\>km\>s$^{-1}$ are larger than those in normal spiral galaxies, and our results are consistent with these results.

\section{Aperture analysis}
\label{sec:aperture_analysis}
\begin{figure}
 \begin{center}
  \includegraphics[width=80mm]{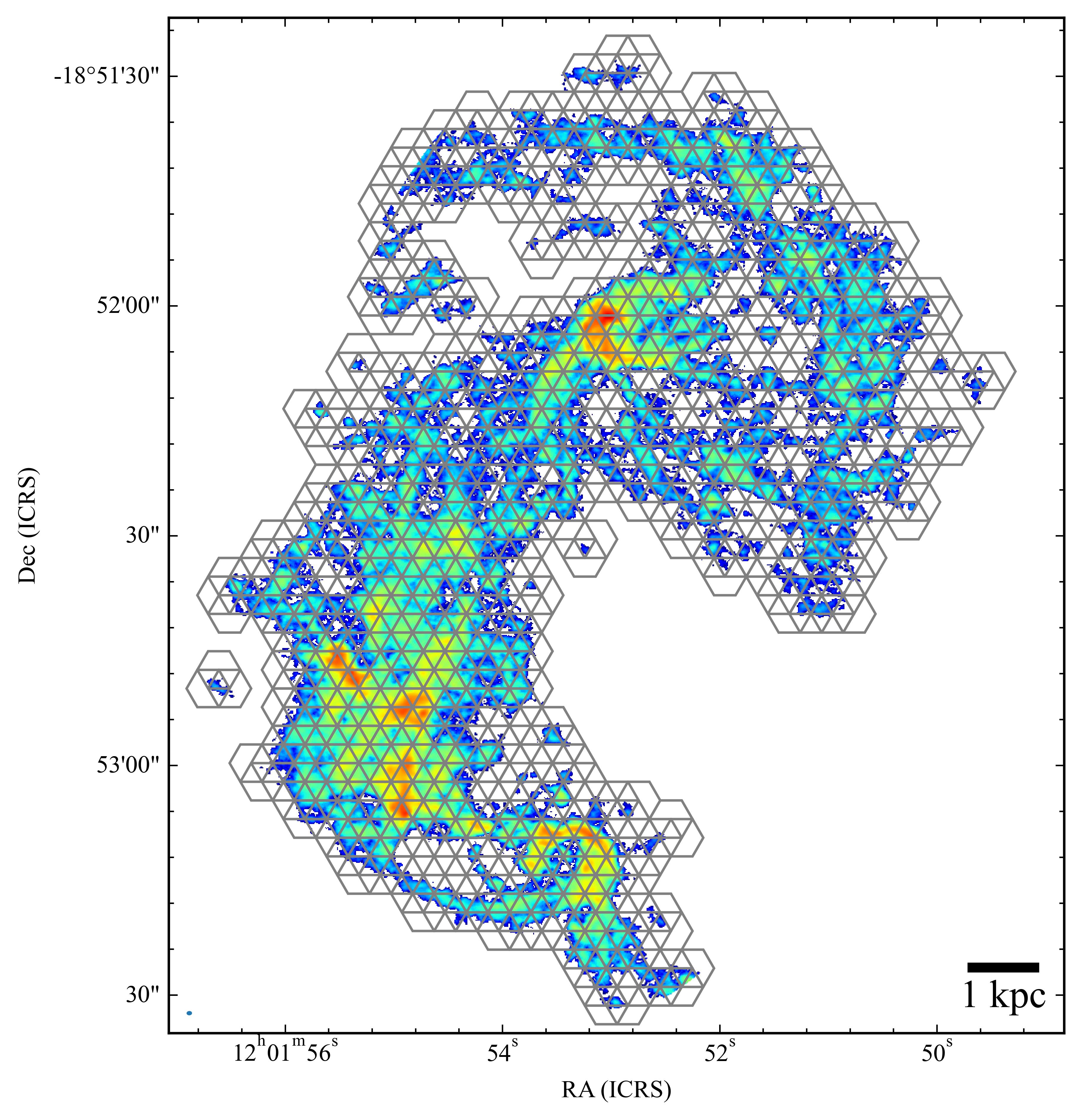}
 \end{center}
\caption{
Hexagon apertures with 300\>pc side superimposed on the velocity integrated CO(1--0) map (color-scale).
The centers of apertures are spaced at an interval of 300 pc.
(See text.)
{Alt text: An image of the central part of the Antennae galaxies with hexagon apertures.}
}
 \label{fig:hex_aper}
\end{figure}

In this study, because the angular resolution of the SFR map is $\sim6''=600$ pc, we took hexagon apertures with 300\>pc side (Figure \ref{fig:hex_aper}) and investigated the relationship among SFR, collision velocity, and typical GMC mass in the apertures. The hexagon apertures are positioned to cover all the GMCs in the FoV. 
The centers of apertures are spaced at an interval of 300 pc representing a form of Nyquist sampling that minimizes arbitrariness in the aperture setting.
The total number of apertures is 876 and the number of the GMCs in an aperture is 1--27 with a typical value of 6.
If we use 200 and 400\>pc as a side of the hexagon of the apertures, the final conclusions are not affected (see the Appendix).

\begin{figure}
 \begin{center}
  \includegraphics[width=80mm]{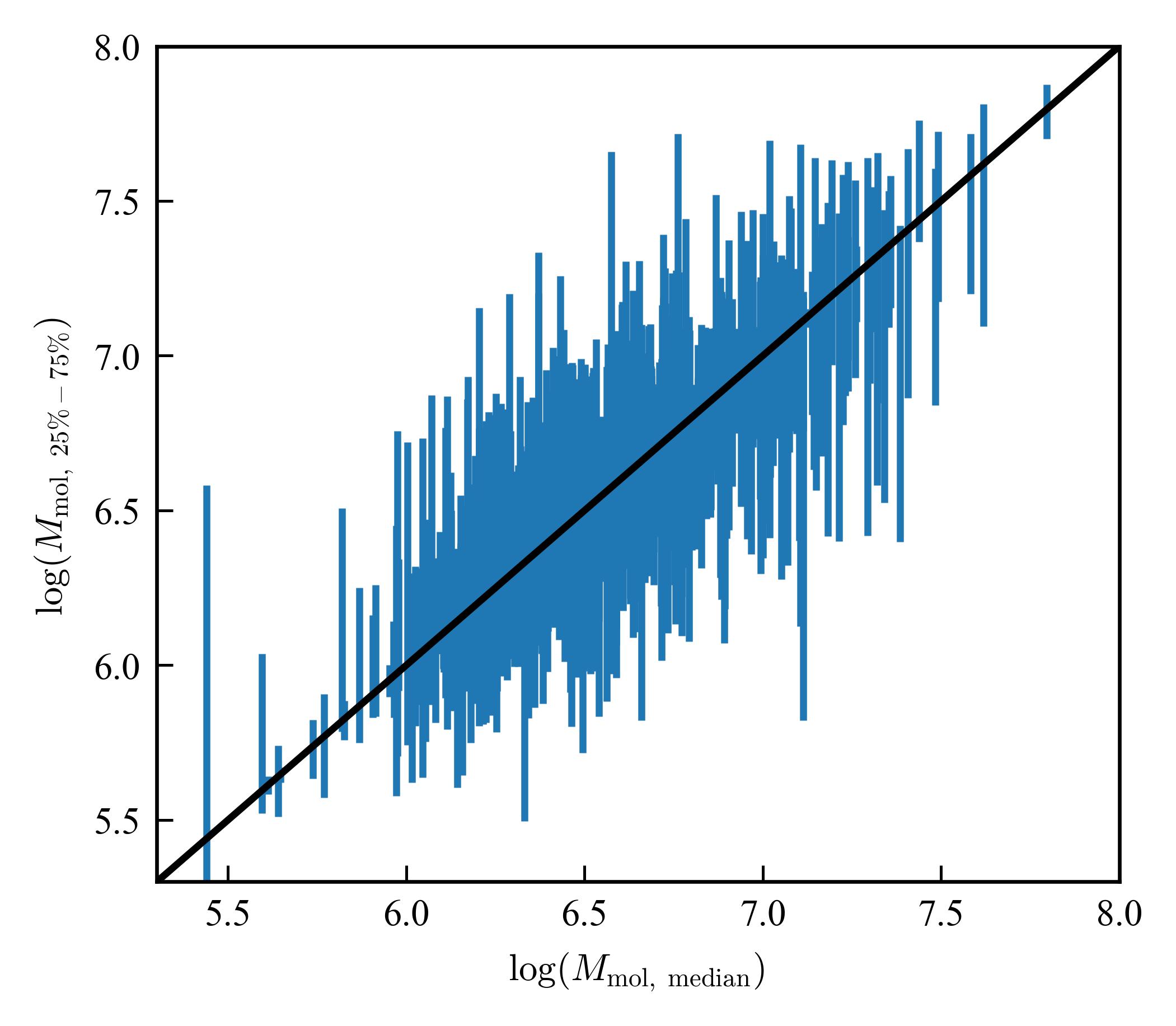}
 \end{center}
\caption{
The relation between median value of the GMC mass and the distribution of GMC mass in each aperture.
The vertical bar represents 25\%--75\% distribution of the GMC mass in each aperture.
The diagonal line is the equal mass line.
{Alt text: A figure with both axis showing the mass range of $10^{5.5}$ to $10^{8.0}$ solar mass.}
}
 \label{fig:M_mol_median_2575}
\end{figure}

\begin{figure*}
 \begin{center}
  \includegraphics[width=160mm]{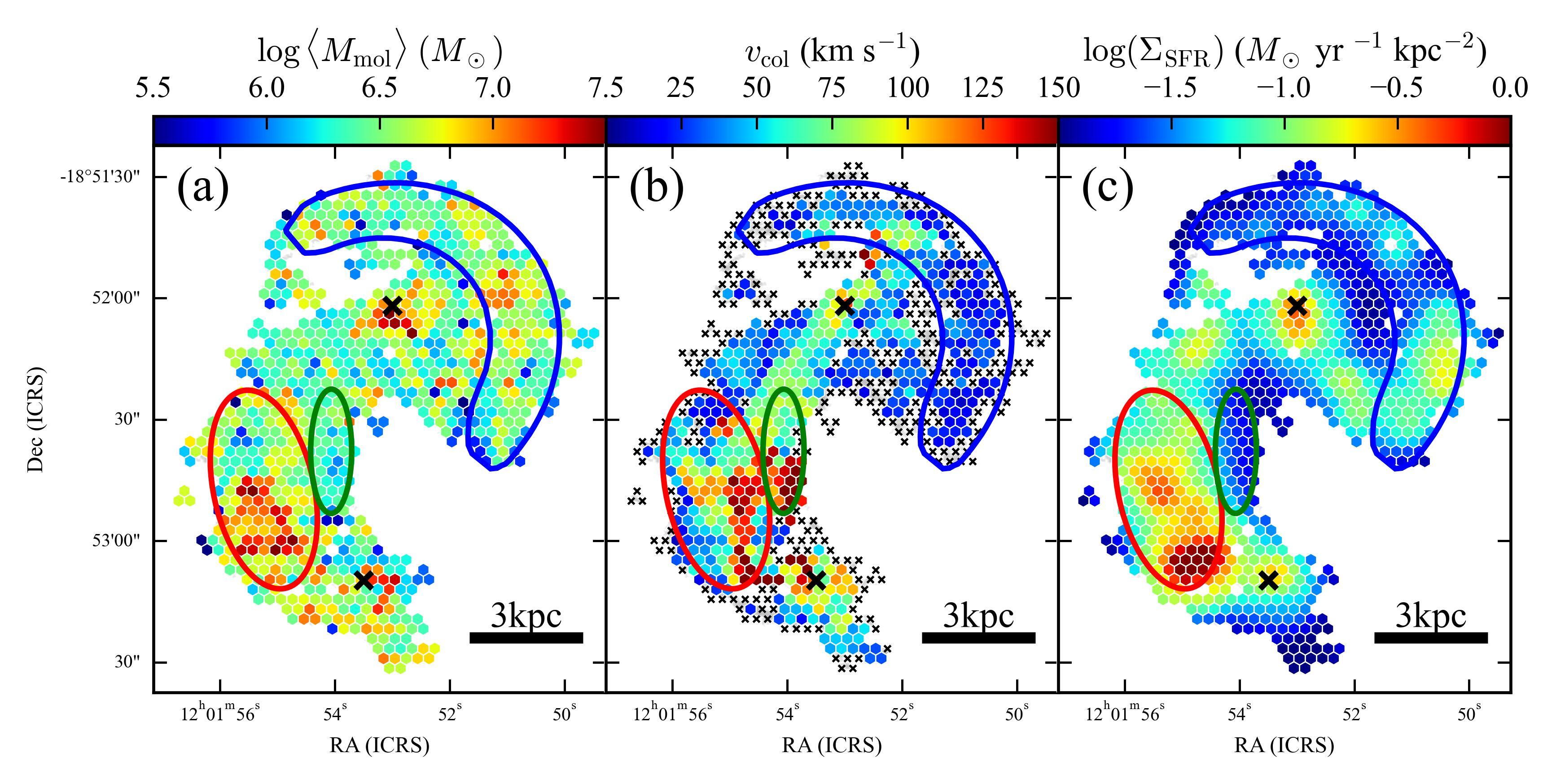}
 \end{center}
\caption{
(a) Spatial distribution of the median GMC mass ($\langle M_{\rm mol} \rangle$), (b) spatial distribution of the collision velocity ($v_{\mathrm{col}}$), and (c) spatial distribution of surface density of SFR ($\Sigma_{\mathrm{
SFR}})$ in the Antennae galaxies.
The $\langle M_{\rm mol} \rangle$, $v_{\mathrm{col}}$, and $\Sigma_{\mathrm{SFR}}$ are derived in each aperture.
The apertures of low collision frequency ($N_{\mathrm{ap}} f_{\mathrm{CCC}}< 0.1$\>Myr$^{-1}$) are indicated with small black crosses in panel (b).
Large black crosses indicate the nuclei of NGC\,4038 and NGC\,4039.
Red ellipse, green ellipse, and blue line are the same as those in figure \ref{fig:Antennae_overview}.
{Alt text: Three aligned images labeled with a to c.
The median GMC mass range is $10^{5.5}$ to $10^{7.5}$ solar mass and the collision velocity range is 20 to 150 kilo meter per second and surface density of star formation rate is 0.01 to 1 Solar mass per year per square kiloparsec.}
}
 \label{fig:vcol_mass_SFR_distribution}
\end{figure*}

For SFR, we used the surface density of SFR in the aperture ($\Sigma_{\mathrm{SFR}}^{\mathrm{ap}}$).
We used the median GMC mass ($\langle M_{\rm mol} \rangle$) in each aperture as the typical mass. 
Figure \ref{fig:M_mol_median_2575} shows the median value versus the 25\%--75\% range of the GMC mass within each aperture.
Most GMC masses are distributed within $\sim0.5$ dex around the median value in the aperture,
showing that the median value well represents the typical mass in each aperture.
Figure \ref{fig:vcol_mass_SFR_distribution}a shows the spatial distribution of the $\langle M_{\rm mol} \rangle$.
In the southern part of the active overlap region where active star formation is seen and nuclear regions, $\langle M_{\rm mol} \rangle$ is large ($>10^7$\>$M_\odot$), 
but in the non-active overlap region, small $\langle M_{\rm mol} \rangle$ ($\lesssim10^6$\>$M_\odot$) is dominant.

\begin{figure}
 \begin{center}
  \includegraphics[width=80mm]{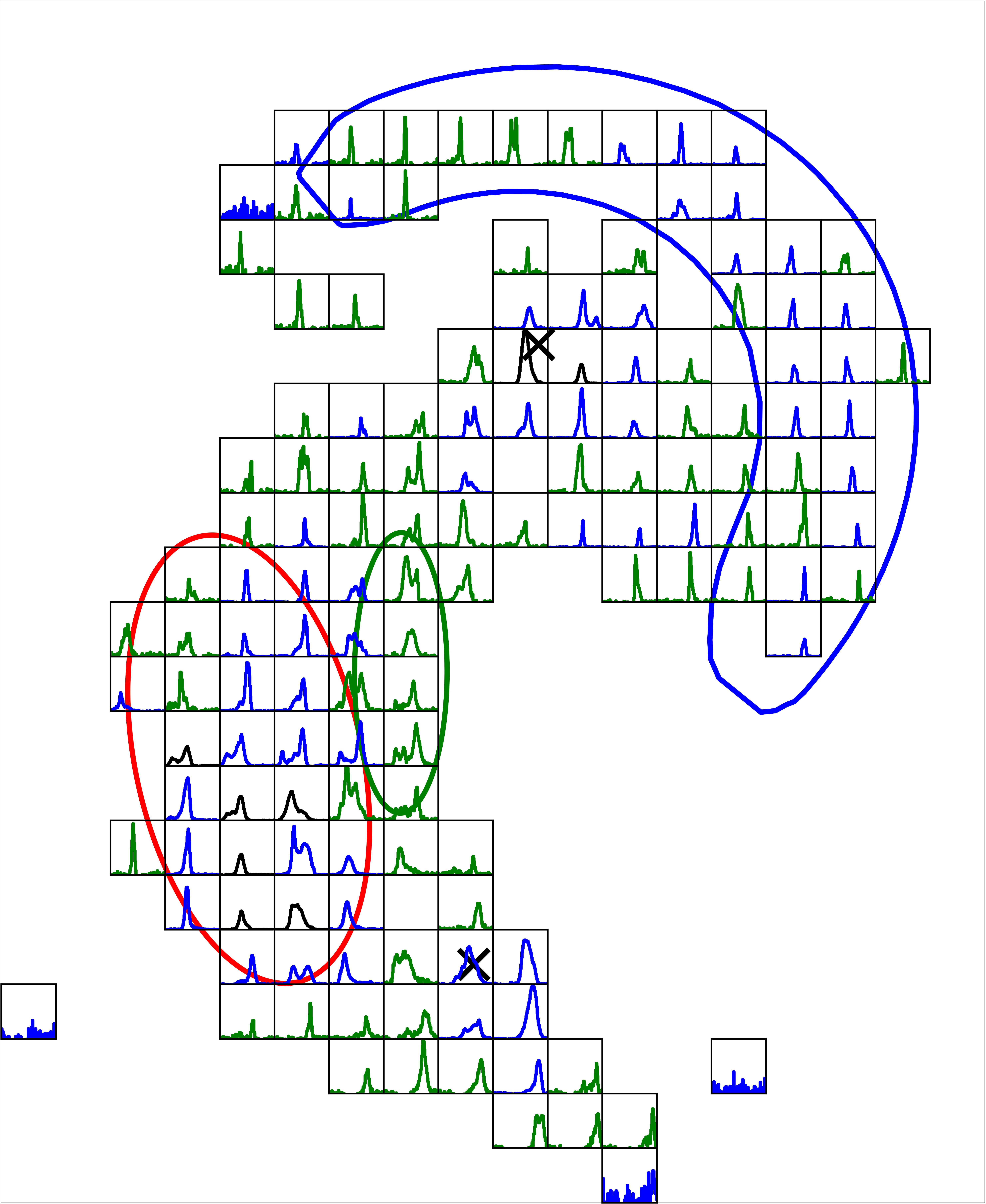}
 \end{center}
\caption{
Map of the CO(1--0) profiles in the Antennae galaxies.
Small boxes show the CO(1--0) profile integrated in the region of $\sim60'' \times 60''$ ($\sim 600\times600$\>pc) square.
The velocity range of 1300--1850\>km\>s$^{-1}$ is shown.
The vertical axes show the brightness temperature.
The black, blue, and green profiles shows the range of 0--5 K, 0--1.7 K, and 0--0.5 K, respectively.
Black crosses indicate the nuclei of NGC\,4038 and NGC\,4039.
Red ellipse, green ellipse, and blue line are the same as those in figure \ref{fig:Antennae_overview}.
{Alt text: An image showing the distribution of the velocity profiles in the Antennae galaxies, where each profiles are placed at each position in the galaxies.}
}
 \label{fig:pixbypix_rebin46}
\end{figure}

When GMCs collide, two or more GMCs with different line-of-sight velocities are expected to exist in a relatively small spatial region.
Therefore, multi-component CO profile such as shown in figure \ref{fig:GMCidentification.example}c indicates the presence of candidates of colliding clouds, 
and the collision velocity of the GMCs can be estimated from the differences of these peaks of the CO profiles.
Figure \ref{fig:pixbypix_rebin46} shows the map of the CO profiles in the Antennae galaxies, 
and multi-component CO profiles are seen throughout the Antennae galaxies, especially in the overlap region (red and green ellipses).
Thus, many colliding GMCs are considered to exist in the Antennae galaxies.

In this study, assuming that GMCs are moving randomly in 3-dimension space in the aperture, we estimated the typical value of the collision velocity of the GMCs in each aperture as
\begin{equation}
v_{\mathrm{col}} = \sqrt{3(\sum_i(v^i_{\mathrm{los}}-\langle v_{\mathrm{los}}\rangle)^2)/N},
\label{eq:collisionvel}
\end{equation}
where $v^i_{\mathrm{los}}$ is the line-of-sight velocity of the $i$-th GMC in the aperture, $\langle v_{\mathrm{los}}\rangle$ is the average value of $v^i_{\mathrm{los}}$ in the aperture, $N$ is the number of the GMCs in the aperture. 
The estimated collision velocity is comparable to the difference of the velocity for the most separated peaks of the CO profile in the aperture.

Figure \ref{fig:vcol_mass_SFR_distribution}b shows the spatial distribution of the collision velocity ($v_{\mathrm{col}}$) calculated from equation (\ref{eq:collisionvel}).
The Antennae galaxies exhibit a wide range of the collision velocity of $\sim10$--$150$\>km\>s$^{-1}$.
Regions with higher collision velocities ($\gtrsim50$\>km\>s$^{-1}$) correspond to the regions with multi-component CO profiles in figure \ref{fig:pixbypix_rebin46} 
and the very large collision velocities ($\gtrsim100$\>km\>s$^{-1}$) are concentrated in the overlap region (red and green ellipses) and the nuclear regions.
\citet{Tsuge+21a, Tsuge+21b} identified a few individual colliding massive clouds with the collision velocity of $\sim100$\>km\>s$^{-1}$ in a part of the active overlap region, and our values are consistent with the value obtained by \citet{Tsuge+21a, Tsuge+21b}.
Figure \ref{fig:vcol_mass_SFR_distribution}c shows surface density of SFR for reference to compare the distributions of $\langle M_{\rm mol} \rangle$, $v_{\mathrm{col}}$, and $\Sigma_{\mathrm{SFR}}^{\mathrm{ap}}$.

In the following sections, after excluding apertures that contain only one GMC, we focus on the apertures where CCC is expected to occur frequently ($N_{\mathrm{ap}} f_{\mathrm{CCC}}>0.1$\>Myr$^{-1}$; discuss in detail in section \ref{sec:discussion}). Out of 876 apertures, 560 apertures meet this criterion.

\section{Results: Relation between cloud-cloud-collisions and SFR in the Antennae galaxies}
\label{sec:results}
\begin{figure*}
 \begin{center}
  \includegraphics[width=160mm]{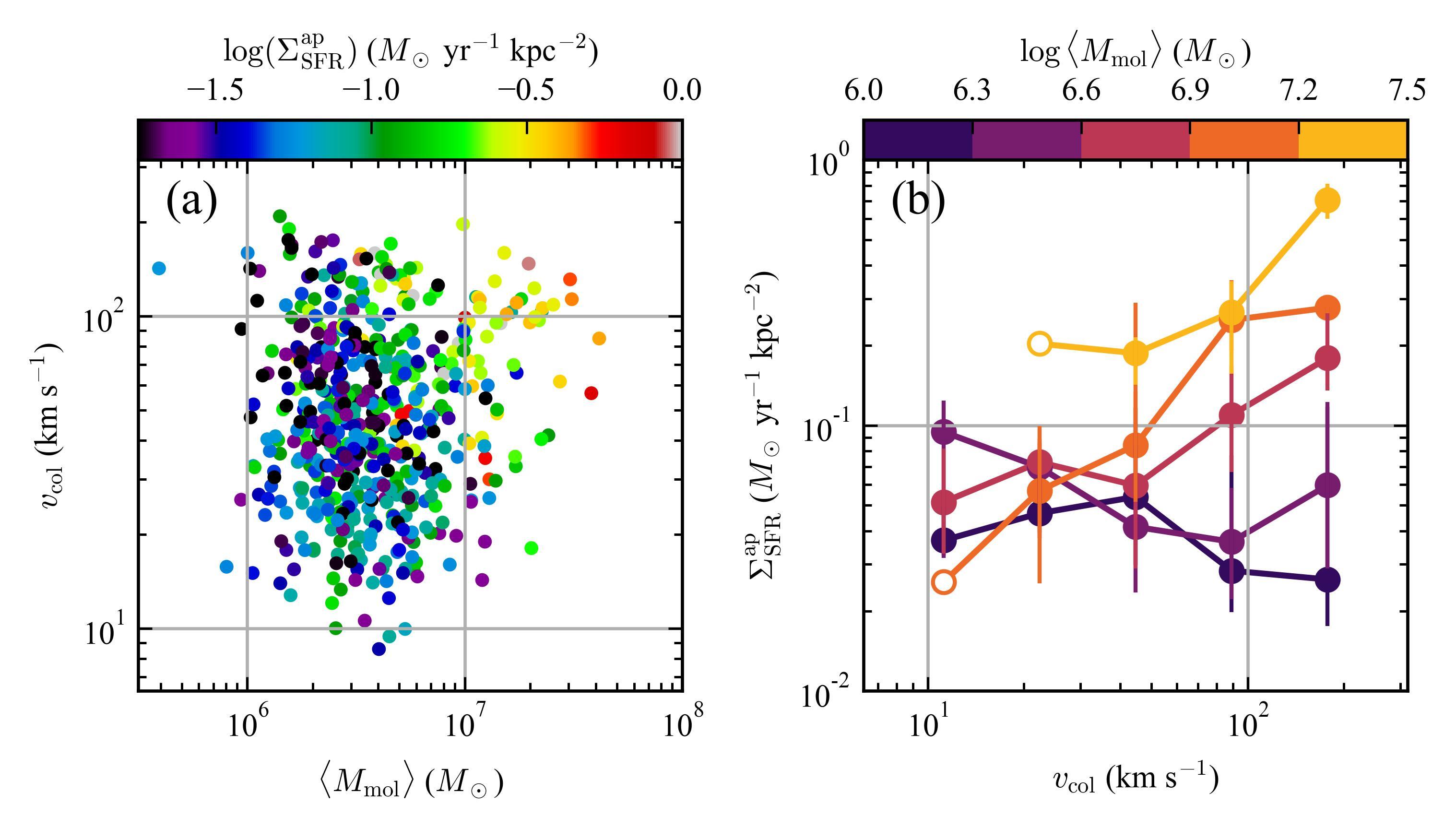}
 \end{center}
\caption{
(a) Surface density of SFR ($\Sigma_{\mathrm{SFR}}^{\mathrm{ap}}$) against the GMC mass and the collision velocity ($v_{\rm col}$) of the GMCs.
The horizontal axis shows the median value of the GMC mass ($\langle M_{\rm mol} \rangle$) and the vertical axis shows $v_{\rm col}$ in the aperture.
The color shows the $\Sigma_{\mathrm{SFR}}^{\mathrm{ap}}$.
(b) $\Sigma_{\mathrm{SFR}}^{\mathrm{ap}}$ against $v_{\rm col}$ by dividing $\langle M_{\rm mol} \rangle$ into 5 bins.
The horizontal axis shows $v_{\rm col}$ and the vertical axis shows $\Sigma_{\mathrm{SFR}}^{\mathrm{ap}}$ in the aperture.
Circles and error bars show the median value and the 25\%--75\% range of the apertures in each mass bin at a collision velocity, respectively.
The color shows $\langle M_{\rm mol} \rangle$.
Open circles indicate that there is only one aperture in a mass and velocity bin.
{Alt text: A scatter plot and a line graph labeled with a and b.}
}
 \label{fig:CCC_SFRplot}
\end{figure*}

Figure \ref{fig:CCC_SFRplot}a shows the dependence of $\Sigma_{\rm SFR}^{\rm ap}$ on $\langle M_{\rm mol} \rangle$ and $v_{\rm col}$.
In this figure, three characteristic parameter spaces are seen:
(i) in the parameter space of high $v_{\rm col}$ of $\sim100$\>km\>s$^{-1}$ and massive ($\langle M_{\rm mol} \rangle \sim10^{7\mathchar`-\mathchar`-8}$\>$M_\odot$) GMCs, the highest $\Sigma_{\mathrm{SFR}}^{\mathrm{ap}}$ of $>10^{-0.4}$\>$M_\odot$\>yr$^{-1}$\>kpc$^{-2}$ is seen,
(ii) in the high $v_{\rm col}$ of $\sim100$\>km\>s$^{-1}$ and relatively low mass ($\langle M_{\rm mol} \rangle \sim10^{6\mathchar`-\mathchar`-7}$\>$M_\odot$) GMCs, 
the lowest $\Sigma_{\mathrm{SFR}}^{\mathrm{ap}}$ of $<10^{-1.3}$\>$M_\odot$\>yr$^{-1}$\>kpc$^{-2}$ and intermediate $\Sigma_{\mathrm{SFR}}^{\mathrm{ap}}$ of 
 $\sim10^{-1.0}$\>$M_\odot$\>yr$^{-1}$\>kpc$^{-2}$ are seen,
and 
(iii) in the lower $v_{\rm col}$ of $\sim10$--$50$\>km\>s$^{-1}$ and relatively low mass ($\sim10^{6\mathchar`-\mathchar`-7}$\>$M_\odot$) GMCs, 
intermediate $\Sigma_{\mathrm{SFR}}^{\mathrm{ap}}$ is seen\footnote{It should be noted that there is a void of the distribution in the right-bottom region in the panel. This is because the collision velocity is considered to be larger than a free-fall velocity between two clouds (see figure 7 by \citet{Maeda+21}).}.

\begin{figure}
 \begin{center}
  \includegraphics[width=80mm]{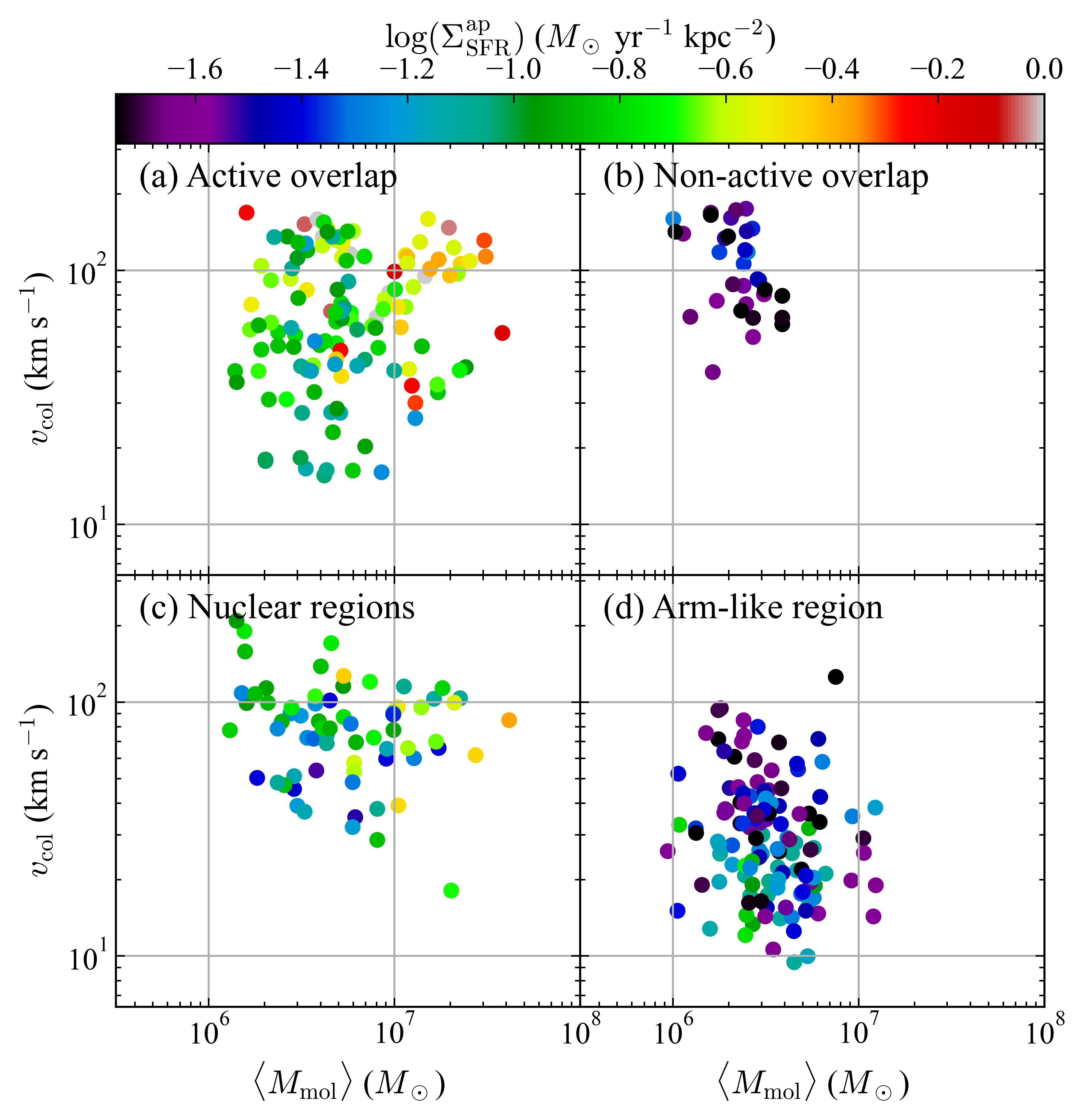}
 \end{center}
\caption{
Surface density of SFR ($\Sigma_{\mathrm{SFR}}^{\mathrm{ap}}$) against the GMC mass and the collision velocity ($v_{\rm col}$) of the GMCs in the active overlap (a), non-active overlap (b), nuclear (c), and arm-like (d) regions.
Horizontal axis, vertical axis, and color are the same as figure \ref{fig:CCC_SFRplot}.
The apertures in the nuclear regions are those within 1\>kpc from the nuclei indicated in figure \ref{fig:Antennae_overview}.
{Alt text: Four scatter plot labeled with a to d.}
}
 \label{fig:CCC_SFRplot_region}
\end{figure}

Figure \ref{fig:CCC_SFRplot_region} shows the dependence of $\Sigma_{\rm SFR}^{\rm ap}$ on $\langle M_{\rm mol} \rangle$ and $v_{\rm col}$ in the four regions: the active overlap (a), non-active overlap (b), nuclear (c), and arm-like (d) regions.
In the active overlap region (panel (a)), data points are located in the parameter space of (i)--(iii) and the data points in the parameter space (i) show the highest $\Sigma_{\mathrm{SFR}}^{\mathrm{ap}}$ of $>10^{-0.4}$\>$M_\odot$\>yr$^{-1}$\>kpc$^{-2}$.
Those data points are located mostly in the southern part of the active overlap region, supporting high-speed collisions of massive GMCs cause active star formation.
In the non-active overlap region (panel (b)) data points with the lowest $\Sigma_{\mathrm{SFR}}^{\mathrm{ap}}$ of $<10^{-1.3}$\>$M_\odot$\>yr$^{-1}$\>kpc$^{-2}$ are concentrated in the parameter space (ii), suggesting that the high-speed collision of low mass GMCs suppresses star formation.
In the nuclear regions (panel (c)), data points are located in the parameter spaces of (i) and (ii), but the data points in the parameter space (i) do not show the highest $\Sigma_{\mathrm{SFR}}^{\mathrm{ap}}$ as seen in the active overlap region.
In the arm-like region (panel (d)), the collision velocity is smaller ($<100$\>km\>s$^{-1}$) than other regions and the trend that $\Sigma_{\mathrm{SFR}}^{\mathrm{ap}}$ decreases with increasing $v_{\rm col}$ is seen.
These results suggest the differences of the SFR across the galactic structures 
link to the differences of the $v_{\rm col}$ and $\langle M_{\rm mol}\rangle$.

To explore this further, figure \ref{fig:CCC_SFRplot}b shows the $\Sigma_{\rm SFR}^{\rm ap}$ as a function of the $v_{\rm col}$, with the $\langle M_{\rm mol} \rangle$ divided into five bins (colored lines). 
Figure \ref{fig:CCC_SFRplot}b shows that if $v_{\rm col}$ is large 
at $\sim100$\>km\>s$^{-1}$, $\Sigma_{\mathrm{SFR}}^{\mathrm{ap}}$ depends on $\langle M_{\rm mol} \rangle$; $\Sigma_{\mathrm{SFR}}^{\mathrm{ap}}$ is higher for the massive GMCs ($\langle M_{\rm mol} \rangle>10^7$\>$M_\odot$), and $\Sigma_{\mathrm{SFR}}^{\mathrm{ap}}$ tends to be lower for the less massive GMCs ($\langle M_{\rm mol} \rangle\sim10^6$\>$M_\odot$).
This suggests that in order to trigger active star formation in a high-speed CCC, large GMC mass is required. 

This trend is qualitatively similar to that seen in the nearby barred spiral galaxies \citep{Maeda+21}.
\citet{Maeda+21} investigated the relationship between the variety of the star formation activities over the galactic environment and the GMC mass and the cloud-collision velocity.
They found that (i) the active star formation is seen in a bar-end region where the collision velocity is large 
and the cloud mass is large,
(ii) the most inactive star formation is seen in a bar region where the collision velocity is large
and the cloud mass is relatively small,
and (iii) the active star-formation is also seen in an arm region where the collision velocity is small
regardless of the GMC mass.


\section{Star formation efficiency for a colliding GMC}
\label{sec:discussion}
While the SFR discussed above is an integrated value within each aperture, this section focuses on the SFR or SFE of a colliding GMC, examining the star formation process at a more fundamental level.
Here, we estimate a SFE of a colliding GMC ($\epsilon_{\mathrm{CCC}}$), defined as the total fraction of GMC gas converted to stars through a CCC, and examine its dependence on the $v_{\rm col}$ and $\langle M_{\rm mol} \rangle$.
Based on the CCC-driven star formation model (e.g., \cite{Tan00}), the SFR integrated within an aperture can be expressed as
\begin{equation}
    \label{eq:eCCC}
    \mathrm{SFR} = \epsilon_{\mathrm{CCC}} \langle M_{\rm mol} \rangle N_{\mathrm{ap}}f_{\mathrm{CCC}},
\end{equation}
where $N_{\mathrm{ap}}$ is the number of the GMCs in the aperture and $f_{\mathrm{CCC}}$ is the collision frequency for one GMC.
From this equation, $\epsilon_{\mathrm{CCC}}$ can be derived.
Strictly speaking, the $\epsilon_{\mathrm{CCC}}$ is the product of the fraction of cloud collisions that successfully lead to star formation ($f_{\mathrm{sf}}$) and the fraction of GMC gas converted into stars during a star-forming collision ($\epsilon$; see \cite{Tan00}). 
However, since these two values cannot be measured separately, in this study we examine only $\epsilon_{\mathrm{CCC}}$.

Assuming that GMCs are in a random motion in 3-dimension, $f_{\mathrm{CCC}}$ within an aperture can be estimated as
\begin{equation}
    \label{eq:fccc}
    f_{\mathrm{CCC}} = \sigma_{\rm c} v_{\rm col} n_{\rm ap}.
\end{equation}
Here, $\sigma_{\mathrm{c}}$ is the collision cross-section, defined as $\pi (2 \langle R\rangle)^2$, where $\langle R\rangle$ is the median GMC radius in the aperture. $n_{\rm ap}$ is the number density of the GMCs in the aperture, defined as $N_{\rm ap}/(Sl)$, where $S$ is the area of the aperture and $l$ is the depth along the line of sight of the volume where the GMCs are moving. We adopted 600\>pc for $l$ to be similar to the size of the apertures.

\begin{figure*}
 \begin{center}
  \includegraphics[width=160mm]{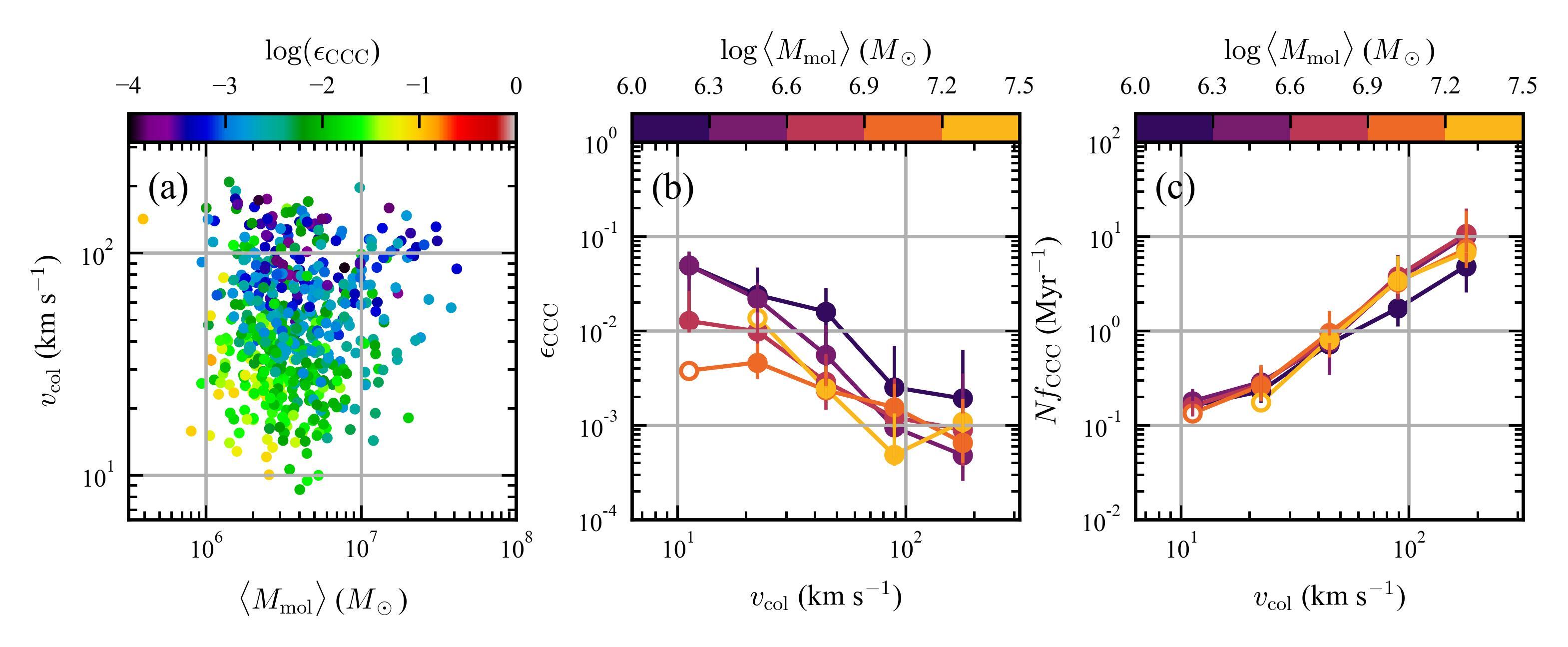}
 \end{center}
\caption{
(a) Star formation efficiency per collision ($\epsilon_{\mathrm{CCC}}$) against the mass ($\langle M_{\mathrm{mol}}\rangle$) and the collision velocity ($v_{\mathrm{col}}$) of the GMCs.
The horizontal and vertical axes are the same as figure \ref{fig:CCC_SFRplot} (a).
The color shows the $\epsilon_{\mathrm{CCC}}$.
(b) $\epsilon_{\mathrm{CCC}}$ against $v_{\mathrm{col}}$ by dividing $\langle M_{\mathrm{mol}}\rangle$ into 5 bins.
The horizontal axis and color are the same as figure \ref{fig:CCC_SFRplot} b.
The vertical axis shows the $\epsilon_{\mathrm{CCC}}$.
(c) Number of collisions per unit time ($N_{\mathrm{ap}} f_{\mathrm{CCC}}$) against $v_{\mathrm{col}}$ by dividing $\langle M_{\mathrm{mol}}\rangle$ into 5 bins.
The horizontal axis and color are the same as figure \ref{fig:CCC_SFRplot} b.
The vertical axis shows the $N_{\mathrm{ap}} f_{\mathrm{CCC}}$.
Open circles in panel (b) and (c) indicate that there is only one aperture in a mass and velocity bin.
{Alt text: A scatter plot and two line graphs with labeled from a to c.}
}
 \label{fig:CCC_eccc_fcccN}
\end{figure*}

\begin{figure}
 \begin{center}
  \includegraphics[width=80mm]{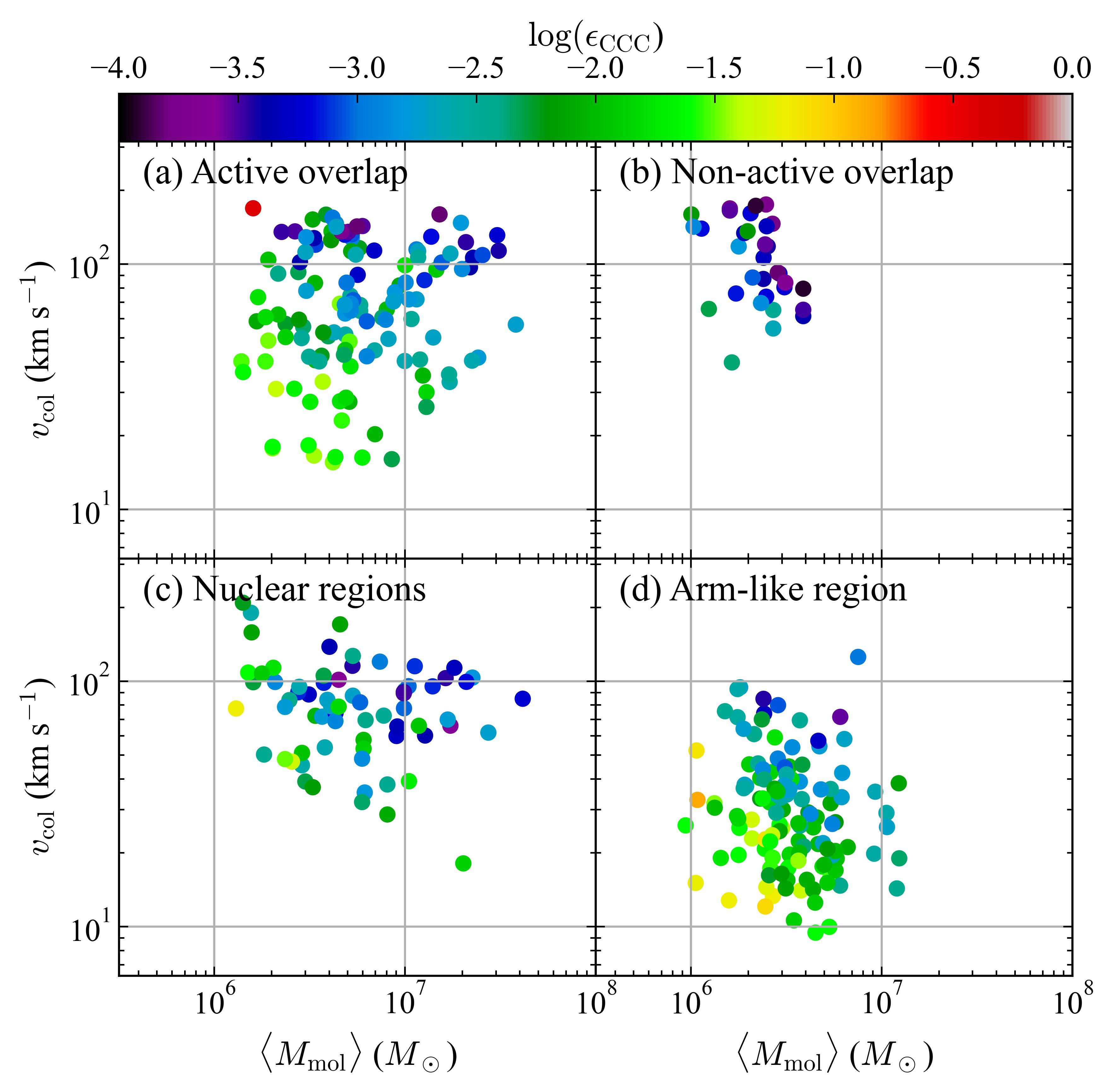}
 \end{center}
\caption{
Star formation efficiency per collision ($\epsilon_{\mathrm{CCC}}$) against the mass ($\langle M_{\mathrm{mol}}\rangle$) and the collision velocity ($v_{\mathrm{col}}$) of the GMCs in the active overlap (a), non-active overlap (b), nuclear (c), and arm-like (d) regions.
Horizontal axis, vertical axis, and color are the same as figure \ref{fig:CCC_eccc_fcccN}.
The apertures in the nuclear regions are the those within 1\>kpc from nuclei indicated in figure \ref{fig:Antennae_overview}.
{Alt text: Four scatter plot labeled with a to d.}
}
 \label{fig:CCC_eccc_fcccN_region}
\end{figure}

Figure \ref{fig:CCC_eccc_fcccN}a 
shows the $\epsilon_{\mathrm{CCC}}$ plotted against the $\langle M_{\rm mol} \rangle$ and $v_{\rm col}$
and figure \ref{fig:CCC_eccc_fcccN_region} shows that in the four regions: The active overlap (a), non-active overlap (b), nuclear (c), and arm-like (d) regions. The same trend that $\epsilon_{\mathrm{CCC}}$ decreases with increasing $v_{\mathrm{col}}$ is seen except for the non-active overlap region where the range of $v_{\mathrm{col}}$ is narrow.
In order to see the trend more clearly, 
figure \ref{fig:CCC_eccc_fcccN}b shows $\epsilon_{\mathrm{CCC}}$ versus $v_{\rm col}$ with $\langle M_{\rm mol} \rangle$ divided into 5 bins.
The $\epsilon_{\mathrm{CCC}}$ is estimated to be $0.001$--$0.03$ (0.1\%--3.0\%).
The $\epsilon_{\mathrm{CCC}}$ decreases as the $v_{\rm col}$ increases, but the $\epsilon_{\mathrm{CCC}}$ is almost the same at a given $v_{\rm col}$, showing a small dependence on $\langle M_{\rm mol} \rangle$;
for data points before the binning, the correlation coefficients for each mass range are $-0.72$ to $-0.45$.
This suggests that the high-speed collision suppresses the star-formation efficiency of the colliding clouds.
A similar trend is also seen in a barred spiral galaxy \citep{Maeda+25}.
 
\citet{Takahira+18} conducted numerical simulations of the CCC and investigated the relationship among the mass of the clouds, collision velocity, and the total mass of formed pre-stellar cores.
They found that the mass fraction of the formed stars to total mass of the colliding clouds ($\epsilon_{\mathrm{CCC}} $) is 0.01--0.1 (1\%--10\%) for colliding clouds with a total mass of $\sim10^{3\mathchar`-\mathchar`-4}$\>$M_\odot$ and collision velocity of $\sim10$--$30$\>km\>s$^{-1}$.
This fraction decreases with increasing collision velocity almost independent of the cloud mass.
They suggested that when the collision velocity is large, a collision duration is too short to form massive pre-stellar cores, resulting in a decrease in $\epsilon_{\mathrm{CCC}} $.
Though 
the masses of the clouds and collision velocities they studied
are smaller than those we studied, 
the dependency of the $\epsilon_{\mathrm{CCC}}$ on the collision velocity and its value is similar to the results obtained in this study,
supporting their suggestion.

Figure \ref{fig:CCC_SFRplot} shows that the most active star formation is seen in the region where the collision velocity is $v_{\mathrm{col}}\sim100$\>km\>s$^{-1}$ and GMC mass is $\langle M_{\mathrm{mol}}\rangle \sim 10^{7\mathchar`-\mathchar`-8}$\>$M_\odot$.
However $\epsilon_{\mathrm{CCC}}$ in this region is almost at the lowest end of the values we obtained.
Then, what causes the most active star formation?
Figure \ref{fig:CCC_eccc_fcccN}c shows the number of collisions per unit time in an aperture ($N_{\mathrm{ap}} f_{\mathrm{CCC}}$) plotted against the $v_{\rm col}$. The $N_{\mathrm{ap}} f_{\mathrm{CCC}}$ increases with increasing $v_{\rm col}$ almost independent of the $\langle M_{\mathrm{mol}}\rangle$.
By comparing the most active star-forming region of $v_{\mathrm{col}}\sim100$\>km\>s$^{-1}$ and $\langle M_{\mathrm{mol}}\rangle \sim 10^{7\mathchar`-\mathchar`-8}$\>$M_\odot$ and inactive star-forming region of $v_{\mathrm{col}}\sim100$\>km\>s$^{-1}$ and $\langle M_{\mathrm{mol}}\rangle \sim 10^{6\mathchar`-\mathchar`-7}$\>$M_\odot$ in 
figure \ref{fig:CCC_SFRplot}a,
both $\epsilon_{\mathrm{CCC}}$ and $N_{\mathrm{ap}} f_{\mathrm{CCC}}$ show little dependence on cloud mass.
Therefore, the difference in the total SFR of the high-speed CCC is considered to be due to the difference in the GMC mass,
because SFR is given by $\epsilon_{\mathrm{CCC}} \langle M_{\rm mol} \rangle N_{\mathrm{ap}}f_{\mathrm{CCC}}$.

\section{Summary}
\label{sec:summary}
Recent studies suggest that the collision velocity of CCC and mass of the GMCs determine the activity of massive star formation.
However, the parameter dependence that have been investigated is mostly limited to the collision velocity of 1--40\>km\>s$^{-1}$.
In order to investigate much faster ($\sim100$\>km\>s$^{-1}$) CCCs, we focused on the Antennae galaxies, which is a nearby colliding galaxies system that shows a wide variety of the star-formation activities.
We derived SFR in the Antennae galaxies using H$\alpha$, mid-infrared, and FUV data with the resolution of $\sim600$\>pc.
Using PYCPROS, we identified GMCs from the ALMA CO(1--0) archival data with the resolution of $\sim70$\>pc that covers a wide area of the central part of the Antennae galaxies.
We set hexagon apertures with 300\>pc side and investigated how SFR on a sub-kpc scale and SFE of a colliding GMC depend on the collision velocity and the GMC mass.
Our main results are as follows:

(a)
Antennae galaxies show a wide range of $\Sigma_{\mathrm{SFR}}\sim0.01$--$1$\>$M_\odot$\>yr$^{-1}$\>kpc$^{-2}$; 
the highest $\Sigma_{\mathrm{SFR}}$ of $\gtrsim 10^{-0.4}$\>$M_{\odot}$\>yr$^{-1}$\>kpc$^{-2}$ is seen in the actively star-forming overlap region, 
intermediate $\Sigma_{\mathrm{SFR}}$ of $\sim0.1$\>$M_{\odot}$\>yr$^{-1}$\>kpc$^{-2}$ is mainly seen in the arm-like region in NGC\,4038, 
and the lowest $\Sigma_{\mathrm{SFR}}$ of $\lesssim10^{-1.3}$\>$M_\odot$\>yr$^{-1}$\>kpc$^{-2}$ is mainly seen in the non-active star-forming overlap region.

(b)
1759 GMCs are identified with the typical properties of $M_{\mathrm{GMC}}\sim 10^{6.5}$\>$M_\odot$, $R\sim60$\>pc, and $\sigma_{v}\sim10$\>km\>s$^{-1}$.

(c)
Cloud collision velocity is estimated from the velocity dispersion among the GMCs in the aperture, assuming the GMCs are moving randomly in 3-dimensional space in it.
The Antennae galaxies show a wide range of the collision velocity of $\sim10$--$150$\>km\>s$^{-1}$.

(d)
The relationship between collision velocity, GMC mass, and star formation activity shows roughly three characteristic parameter spaces. 
(i) high-speed collisions ($\sim100$\>km\>s$^{-1}$) of massive ($\sim10^{7\mathchar`-\mathchar`-8}$\>$M_\odot$) GMCs show the highest $\Sigma_{\mathrm{SFR}}^{\mathrm{ap}}$,
(ii) high-speed collisions ($\sim100$\>km\>s$^{-1}$) of low mass ($\sim10^{6\mathchar`-\mathchar`-7}$\>$M_\odot$) GMCs show the lowest and intermediate $\Sigma_{\mathrm{SFR}}^{\mathrm{ap}}$, and
(iii) low-speed collisions ($\sim10$--$50$\>km\>s$^{-1}$) of low mass ($\sim10^{6\mathchar`-\mathchar`-7}$\>$M_\odot$) GMCs show intermediate $\Sigma_{\mathrm{SFR}}^{\mathrm{ap}}$.
The parameter spaces link to the variations of the star formation activities over the galactic structures:
CCCs in parameter space (i) are mainly seen in the active overlap region and the nuclear regions, 
CCCs in parameter space (ii) are mainly seen in the non-active overlap region,
and 
CCCs in parameter space (iii) are mainly seen in the other regions including the arm-like region.

(e)
Star formation efficiency per collision ($\epsilon_{\mathrm{CCC}}= \mathrm{SFR}/ (\langle M_{\mathrm{mol}} \rangle N_{\mathrm{ap}}f_{\mathrm{CCC}})$) is estimated to be $0.001$--$0.03$ (0.1\%--3.0\%) and decreases with increasing collision velocity almost independent of the cloud mass; 
even in the parameter space of (i) that shows the highest $\Sigma_{\mathrm{SFR}}^{\mathrm{ap}}$, $\epsilon_{\mathrm{CCC}}$ is the smallest in the Antennae galaxies. 
The most active star formation is consider to come from the large mass of the cloud.

This study statistically clarified the connection between the collision velocity and GMC mass of high-speed CCC and star-formation activity in the colliding galaxy, the Antennae galaxies.
However, we only focused on the properties in sub-kpc regions and it is desirable to identify individual colliding clouds using more sensitive and high-resolution data to understand more detailed properties of CCC and massive star formation.
Future observations of other nearby colliding galaxies are also important.

\bigskip
\begin{ack}
We thank the referee for valuable comments that improved the paper.
We are grateful to R. Yamada and F. Demachi for their comments on GMC identification.
K.O. is supported by JSPS KAKENHI grant No. JP23K03458.
F.M. is supported by JSPS KAKENHI grant No. JP23K13142.
This paper makes use of the following ALMA data: ADS/JAO.ALMA2018.1.00272. ALMA is a partnership of ESO (representing its member states), NSF (USA) and NINS (Japan), together with NRC (Canada), NSTC and ASIAA (Taiwan), and KASI (Republic of Korea), in cooperation with the Republic of Chile. The Joint ALMA Observatory is operated by ESO, AUI/NRAO and NAOJ. We thank the East Asian ALMA Regional Center for providing calibrated visibility data. 
Some of the data presented in this paper were obtained from the Mikulski Archive for Space Telescopes (MAST) at the Space Telescope Science Institute. The specific observations analyzed can be accessed via \url{http://dx.doi.org/10.17909/4nhk-mm66}. STScI is operated by the Association of Universities for Research in Astronomy, Inc., under NASA contract NAS5-26555. Support to MAST for these data is provided by the NASA Office of Space Science via grant NAG5-7584 and by other grants and contracts.
This work is based in part on observations made with the Spitzer Space Telescope, which was operated by the Jet Propulsion Laboratory, California Institute of Technology under a contract with NASA.
This work is based in part on observations made with the Herschel Space Observatory, Herschel is an ESA space observatory with science instruments provided by European-led Principal Investigator consortia and with important participation from NASA.
This publication makes use of data products from the Two Micron All Sky Survey, which is a joint project of the University of Massachusetts and the Infrared Processing and Analysis Center/California Institute of Technology, funded by the National Aeronautics and Space Administration and the National Science Foundation. 

\end{ack}

\appendix 
\section{Effects of Parameter Choices}
\label{sec:appendix}
Choice of parameters of the PYCPROPS \citep{Rosolowsky+21} affects the properties of the GMC identified (Section \ref{sec:GMCidentification}).
Here, we examine the effects of parameters $\delta$ and compactness, which may affect the properties of the GMCs rather significantly, and discuss the robustness of the conclusions.
The aperture size adopted in section \ref{sec:aperture_analysis} may also affect the estimations of the typical mass and the collision velocity, and we also examine its effect.

\subsection{Delta ($\delta$)}
\begin{figure*}
 \begin{center}
  \includegraphics[width=160mm]{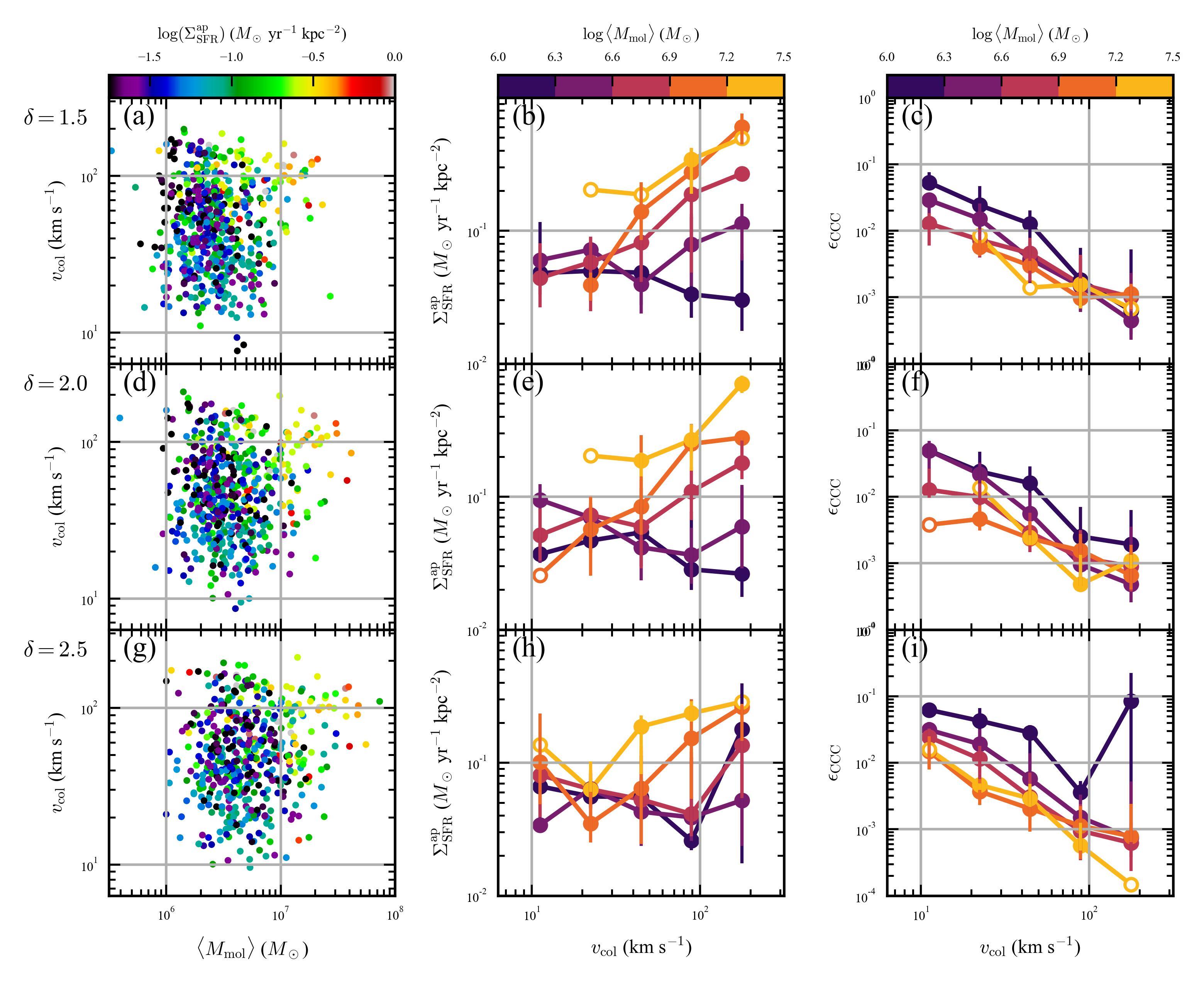}
 \end{center}
\caption{
Same as figure \ref{fig:CCC_SFRplot}a, \ref{fig:CCC_SFRplot}b, and figure\ref{fig:CCC_eccc_fcccN}b,
but for $\delta = 1.5, 2.0$, and $2.5$ for (a)--(c), (d)--(f), and (g)--(i), respectively.
{Alt text: Three scatter plots labeled with a, d and g and six line graphs labeled with b, c, e, f, h and i.}
}
 \label{fig:robust.delta}
\end{figure*}

If $\delta$ is large, a local maximum with a small value is not separated from a massive cloud, then the number of the GMCs identified decreases and the sizes and masses of the GMCs increase.
If we use $\delta=$1.5 and 2.5 instead of $\delta=2.0$, 2243 and 1415 GMCs are identified, respectively.
Figure \ref{fig:robust.delta} shows the same figures as figure \ref{fig:CCC_SFRplot}a, \ref{fig:CCC_SFRplot}b, and figure \ref{fig:CCC_eccc_fcccN}b but for $\delta =1.5, 2.0, $ and $2.5$.
Figures \ref{fig:robust.delta}a, \ref{fig:robust.delta}d, and \ref{fig:robust.delta}g show that the median value of the cloud mass increases with increasing value of $\delta$, resulting in the slight shift of the distribution to the right,
however, three characteristic parameter spaces discussed in section \ref{sec:results} remain.
$\Sigma_{\mathrm{SFR}}^{\mathrm{ap}}$ increases with increasing cloud mass at the large collision velocity of $\gtrsim100$\>km\>s$^{-1}$.
In figures \ref{fig:robust.delta}c, \ref{fig:robust.delta}f, and \ref{fig:robust.delta}i, $\epsilon_{\mathrm{CCC}}$ decreases with increasing collision velocity without depending on the GMC mass; the tendency does not change so much.

\subsection{Compactness}
\begin{figure*}
 \begin{center}
  \includegraphics[width=160mm]{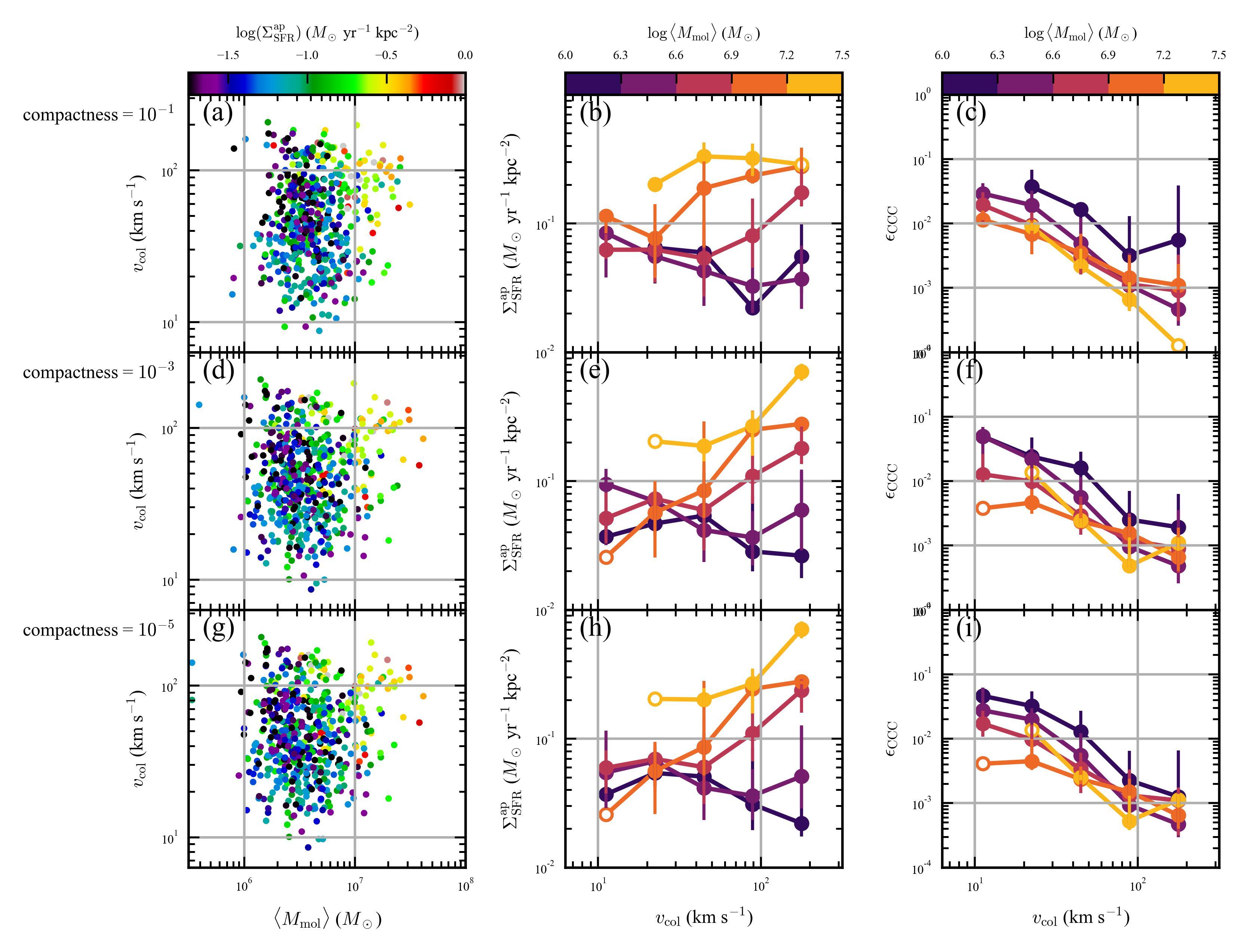}
 \end{center}
\caption{
Same as figure \ref{fig:CCC_SFRplot}a, \ref{fig:CCC_SFRplot}b, and figure\ref{fig:CCC_eccc_fcccN}b,
but for compactness$ = 0.1, 0.001$, and $0.00001$ for (a)--(c), (d)--(f), and (g)--(i), respectively.
{Alt text: Three scatter plots labeled with a, d and g and six line graphs labeled with b, c, e, f, h and i.}
}
 \label{fig:robust.compactness}
\end{figure*}

The compactness parameter affects the boundary of a GMC and affects the mass and the size of the GMC.
In this study, we used 0.001 for compactness parameter to obtain natural boundaries.
If we use larger values ($>0.1$), masses and sizes of 25\% of the low mass clouds ($\lesssim10^5$\>$M_\odot$) change by an order of magnitude.
However, the final conclusions are not affected so much.
Figure \ref{fig:robust.compactness} shows the same figures as figure \ref{fig:CCC_SFRplot}a, \ref{fig:CCC_SFRplot}b, and figure\ref{fig:CCC_eccc_fcccN}b but for compactness $ = 0.1, 0.001$, and $0.00001$.
For compactness = 0.1, though figures \ref{fig:robust.compactness}a--\ref{fig:robust.compactness}c show a subtle change compared with figures \ref{fig:robust.compactness}d--\ref{fig:robust.compactness}f (compactness = 0.001), the overall trends do not change.
For compactness = 0.00001, figures \ref{fig:robust.compactness}g--\ref{fig:robust.compactness}i show no clear change compared with the figures for compactness = 0.001.

\subsection{Aperture size}
The aperture size affects the typical mass and the collision velocity in the aperture.
Taking a very small aperture may contain only one GMC (or even no GMC) and does not make sense, while taking a very large aperture size includes a velocity field in the galaxy, and the collision velocity estimation does not give an appropriate value.
Thus, we examine the cases with a slightly smaller value (200\>pc) and a larger value (400\>pc) as a side of the hexagon apertures. 
The final conclusions are not affected so much as seen below.
Figure \ref{fig:robust.aperture_size} shows the same figures as figure \ref{fig:CCC_SFRplot}a, \ref{fig:CCC_SFRplot}b, and figure \ref{fig:CCC_eccc_fcccN}b but for the aperture size of 200\>pc, 300\>pc, and 400\>pc.
Figures \ref{fig:robust.aperture_size}a, \ref{fig:robust.aperture_size}d, and \ref{fig:robust.aperture_size}g show that though the distribution of the data points becomes more compact when the aperture size increases, the overall trend does not change.
Figures \ref{fig:robust.aperture_size}b, \ref{fig:robust.aperture_size}e, and \ref{fig:robust.aperture_size}h, and \ref{fig:robust.aperture_size}c, \ref{fig:robust.aperture_size}f, and \ref{fig:robust.aperture_size}i show similar trends.
 
\begin{figure*}
 \begin{center}
  \includegraphics[width=160mm]{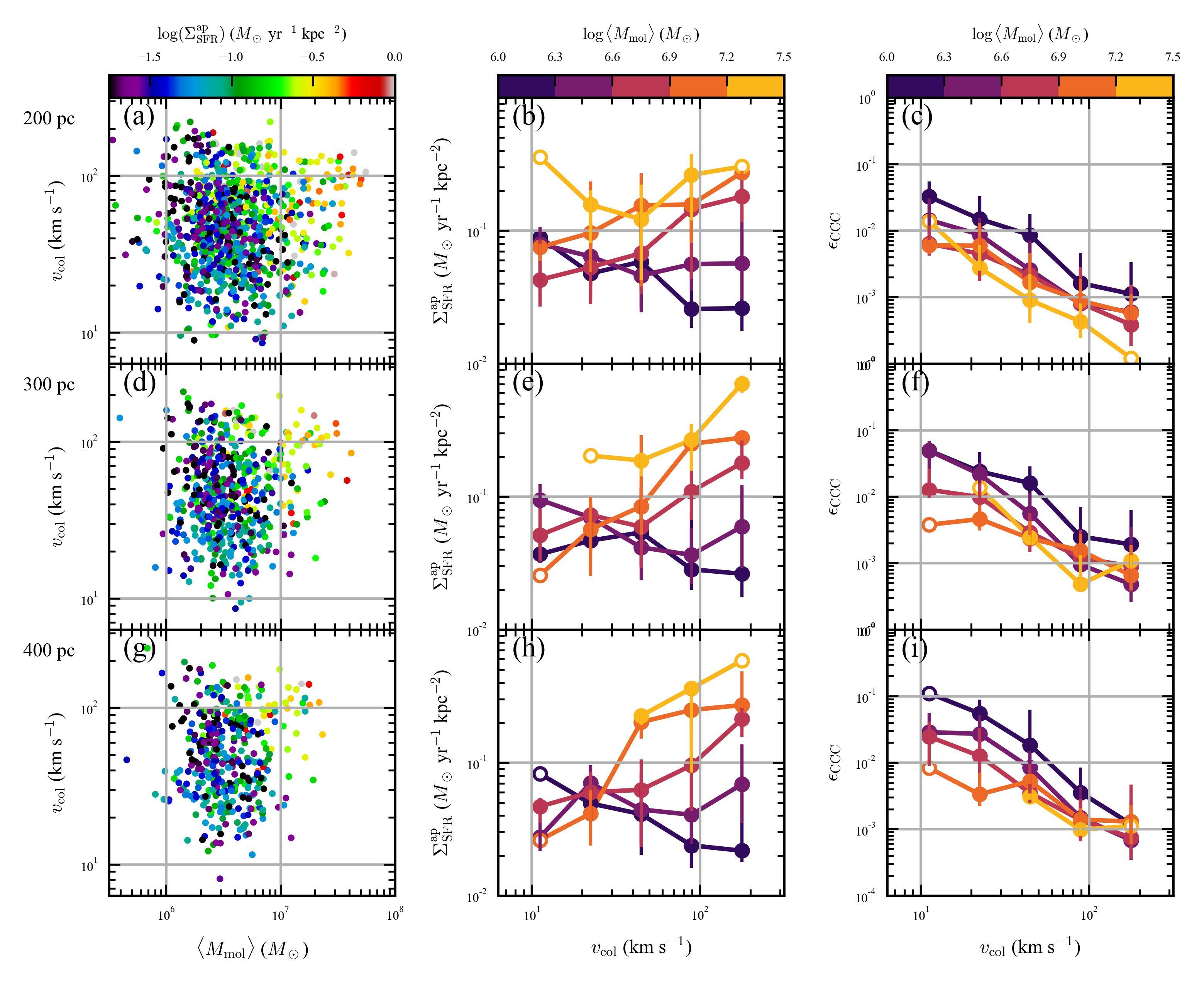}
 \end{center}
\caption{
Same as figure \ref{fig:CCC_SFRplot}a, \ref{fig:CCC_SFRplot}b, and figure\ref{fig:CCC_eccc_fcccN}b,
but for a side of the hexagon apertures of 200\>pc, 300\>pc, and 400\>pc for (a)--(c), (d)--(f), and (g)--(i), respectively.
{Alt text: Three scatter plots labeled with a, d and g and six line graphs labeled with b, c, e, f, h and i.}
}
 \label{fig:robust.aperture_size}
\end{figure*}


\end{document}